\documentclass[aip,jcp,reprint,amsmath,amssymb]{revtex4-2}
\usepackage{graphicx}
\usepackage{bm}
\usepackage{braket}
\usepackage{algorithm}
\usepackage{algpseudocode}

\usepackage{subcaption}
\usepackage{enumitem}
\usepackage[section]{placeins}
\usepackage[dvipsnames]{xcolor}
\usepackage{hyperref}
\hypersetup{hidelinks}
\usepackage{url}
\usepackage{xr-hyper}
\definecolor{darkblue}{rgb}{0,0,.65}
\definecolor{darkgreen}{rgb}{0.28,0.41,0.19}

\makeatletter
\AtBeginDocument{%
  \long\def\@makecaption#1#2{%
    \vskip\abovecaptionskip
    {\leftskip\z@ \rightskip\z@ \parfillskip\z@ plus 1fil\relax
     #1: #2\par}%
    \vskip\belowcaptionskip}%
}
\makeatother

\begin{document}

\title{Interpolative Separable Density-Fitting for Transcorrelated Hamiltonians}
\author{Ke Liao}
\email{ke.liao.whu@gmail.com}
\affiliation{Max Planck Institute for Solid State Research, Heisenbergstrasse 1, 70569 Stuttgart, Germany}
\affiliation{Department of Chemistry, Yale University, Sterling Chemistry Laboratory, 225 Prospect St, New Haven, CT 06511, USA}
\author{Yifan Cheng}
\affiliation{Max Planck Institute for Solid State Research, Heisenbergstrasse 1, 70569 Stuttgart, Germany}
\author{Werner Dobrautz}
\affiliation{Center for Advanced Systems Understanding (CASUS), Helmholtz-Zentrum Dresden-Rossendorf (HZDR), 02826 G\"orlitz, Germany}
\affiliation{Center for Scalable Data Analytics and Artificial Intelligence (ScaDS.AI), TU Dresden, 01062 Dresden, Germany}
\author{Tianyu Zhu}
\email{tianyu.zhu@yale.edu}
\affiliation{Department of Chemistry, Yale University, Sterling Chemistry Laboratory, 225 Prospect St, New Haven, CT 06511, USA}
\author{Ali Alavi}
\email{a.alavi@fkf.mpg.de}
\affiliation{Max Planck Institute for Solid State Research, Heisenbergstrasse 1, 70569 Stuttgart, Germany}
\date{\today}

\begin{abstract}
    The transcorrelated (TC) method folds a Jastrow correlator into the
    Hamiltonian by similarity transformation, building the electron--electron
    cusp into the effective interaction and thereby dramatically accelerating
    the convergence of correlated calculations toward the complete-basis-set
    (CBS) limit. We make this framework practical for large systems and
    flexible, multi-center correlators by compressing the grid-evaluated TC
    integrals with the interpolative separable density-fitting (ISDF)
    approximation, combined with the effective two-body (xTC) treatment of the
    three-body operator. This low-rank representation reduces storage and
    integration costs by orders of magnitude, and a multi-GPU implementation
    with automatic differentiation of the correlator makes the construction
    routine for large basis sets. We demonstrate the resulting ISDF-xTC-CCSD
    method on the linear hydrogen chain, reaching the joint thermodynamic and
    CBS limits with basis sets up to cc-pV5Z in agreement with
    state-of-the-art many-body references to within about 1~mHa/atom, and on
    the benzene ground-state energy with up to 1200 orbitals (cc-pCV5Z),
    where the method attains state-of-the-art accuracy at the
    coupled cluster singles and doubles level and its CBS extrapolation is
    markedly more robust than that of conventional coupled-cluster methods.
\end{abstract}

\maketitle

\section{Introduction}

The accurate treatment of electron correlation is the central challenge of
\emph{ab initio} electronic structure theory, and a major practical obstacle is
the slow convergence of correlated energies with the size of the one-particle
basis. Because an antisymmetric product of smooth orbitals cannot reproduce the
cusp in the wavefunction as two electrons coalesce, correlation energies
converge only as $\mathcal{O}(X^{-3})$ in the cardinal number $X$ of a
correlation-consistent basis set\cite{kutzelnigg1992,helgaker1997}, so that reaching the complete-basis-set (CBS)
limit with orbital-product expansions alone demands prohibitively large bases.

A well-established route around this bottleneck is to build the
electron--electron cusp directly into the wavefunction. Explicitly correlated
R12 and F12 methods augment the orbital expansion with terms that depend
explicitly on the interelectronic distance $r_{12}$, restoring the cusp and
accelerating basis-set convergence by roughly two cardinal
numbers\cite{kutzelnigg1985,klopper1987,hattig2012}. These methods
deliver near-CBS correlation energies in modest bases, but require specialized
many-electron integrals and are tailored to particular correlation-method
implementations. A complementary strategy, pursued here, folds the correlation
factor into the Hamiltonian itself rather than the wavefunction.

The transcorrelated (TC) method, originally introduced by Boys and
Handy\cite{boys1969b,boys1969c,boys1969d,boys1969e,boys1969}, realizes this idea. Similarity
transforming the Hamiltonian with a Jastrow factor,
$\bar H = F^{-1}\hat H F$ with $F = e^{\hat\tau}$, builds the electron--electron
cusp and other correlation effects directly into the Hamiltonian, markedly accelerating basis-set
convergence and extending the reach of low-order coupled-cluster
approximations into strongly correlated regimes. The price is that the
transcorrelated Hamiltonian acquires, beyond the usual two-body interaction, an
additional effective two-body term and a genuine three-body operator, whose
explicit (rank-six) treatment is prohibitive beyond small systems. Interest in
the method has grown steadily across several groups and correlation
treatments. Tsuneyuki and co-workers pioneered its application to
solids, coupling the correlator optimization to variational Monte Carlo and
developing efficient plane-wave algorithms and correlated band
structures\cite{umezawa2003,sakuma2006,ochi2012,ochi2017}. Alavi and
co-workers combined the TC method with full configuration-interaction quantum
Monte Carlo\cite{luo2018a,haupt2025}, coupled-cluster
theory\cite{liao2021e,schraivogel2021,schraivogel2023}, and the density-matrix
renormalization group\cite{liao2023}, developed systematic Jastrow-factor
optimizations\cite{haupt2023,filip2025jastrow}, and introduced the effective
transcorrelated (xTC) scheme\cite{liao2021e,christlmaier2023,kats2024}, which
removes the three-body bottleneck by contracting the three-body operator with
the reference one-particle density matrix into an effective two-body
correction, restoring the formal scaling of a conventional two-body method.
Reiher and co-workers formulated transcorrelated matrix-product operators and
non-iterative triples corrections\cite{baiardi2022,morchen2025}; Ten-no
introduced a nonunitary projective transcorrelation inspired by the F12
ansatz\cite{tenno2023}; and Giner and co-workers constructed inexpensive
three-body correlation factors with analytic
integrals\cite{giner2021,ammar2023}. These developments have been applied to
atoms and small molecules\cite{filip2025}, the uniform electron gas and
periodic solids\cite{luo2018a,liao2021e,simula2025solids}, and even
quantum-computing hardware\cite{dobrautz2024}---but almost always for
relatively small systems.

Scaling the transcorrelated method to larger systems exposes a second, less widely
appreciated bottleneck: the evaluation of the
transcorrelated two-body integrals themselves. For simple correlators these can
be obtained analytically\cite{giner2021,ammar2023,baiardi2022}, but analytic
schemes are correlator-specific and
become cumbersome---or intractable---for the flexible, multi-center correlators,
such as the Boys--Handy\cite{boys1969} or Drummond--Towler--Needs\cite{drummond2004}
forms, that are needed to capture
electron--electron, electron--nucleus, and three-body
electron--electron--nucleus correlation simultaneously. Numerical real-space
integration is fully general and correlator-agnostic, but its naive cost is
prohibitive: evaluating the transcorrelated two-body integrals requires a
double sum over grid points for every orbital pair, scaling as
$\mathcal{O}(N_g^2 N_{\rm orb}^2)$ in the number of grid points $N_g$;
explicit evaluation of the full three-body (rank-six) integrals is more
expensive still; and even after the xTC contraction eliminates the explicit
three-body tensor, the assembly of the effective two-body correction---the
dominant cost in practice---couples the same double grid sums to
$\mathcal{O}(N_g N_{\rm orb}^4)$ orbital-pair contractions. None of these
costs is affordable for production calculations without further
factorization.

We address this bottleneck with the interpolative separable density-fitting
(ISDF) approximation\cite{lu2015,hu2017,qin2023isdf}, which compresses products of orbitals
(and orbital gradients) on a real-space grid into a low-rank representation
anchored at a small set of interpolation points, and which has accelerated
density fitting\cite{lu2015,qin2023isdf}, hybrid density-functional theory and
exact exchange in molecules and solids\cite{hu2017,dong2018,wu2022,rettig2023},
$GW$ and random-phase-approximation
calculations\cite{gao2020,yeh2023,yeh2024}, post-Hartree--Fock
methods\cite{hohenstein2012,lee2020thc,malone2019,malone2020,luo2025}, quantum
embedding and local correlation in crystalline materials\cite{yang2026}, and
compact Hamiltonian representations for quantum computing\cite{lee2021qc}. We show that the transcorrelated two-body ($K$) and effective two-body
(xTC, $\Delta U$) integrals inherit this low-rank structure, and we build them
for a general differentiable Jastrow factor whose gradients are obtained by automatic
differentiation, eliminating the need for hand-coded, correlator-specific
integral routines.

This work contributes: (i) an ISDF factorization of the
transcorrelated $K$ and $\Delta U$ integrals for a general grid-evaluated
correlator, with a pivoted-Cholesky selection of interpolation points that is
independent of the Jastrow; (ii) a GPU-compatible Jastrow-optimization workflow
that combines the reference-variance VMC objective of Ref.~\onlinecite{haupt2023}---complementary
to recent deterministic Jastrow-optimization schemes\cite{filip2025jastrow}---with
automatic-differentiation gradients and Polyak--Ruppert parameter averaging;
(iii) a GPU-accelerated producer/consumer pipeline with multi-GPU dispatch support
and out-of-core storage that accelerates both the ISDF integral construction and
the downstream CCSD solver, keeping the approach tractable for up to 1200
orbitals in large basis sets; and (iv) integration with a restricted CCSD solver
for non-Hermitian transcorrelated integrals, building on earlier
transcorrelated coupled-cluster work\cite{liao2021e,schraivogel2021,schraivogel2023}. We validate
the ISDF approximation on two
benchmarks: the linear hydrogen chain, where transcorrelation accelerates
basis-set convergence and the thermodynamic-limit energies agree with
state-of-the-art AFQMC+$\Delta$DMRG references\cite{motta2017} to within
$1$~mHa/atom, and the benzene molecule, where the construction remains
tractable up to the cc-pCV5Z basis ($N_{\rm orb}=1200$).

\section{Theoretical Framework}

\subsection{Transcorrelation theory}

The transcorrelated Hamiltonian is obtained by a similarity transformation of the
bare electronic Hamiltonian $\hat{H}$ in first quantization:
\begin{equation}
    \bar{H} = F^{-1}\,\hat{H}\,F,
    \qquad F = e^{\hat{\tau}},
\end{equation}
where $F$ is the Jastrow factor and $\hat{\tau}$ is the Jastrow operator. The
production correlator contains a one-electron nuclear-cusp term $\chi({\bf r})$~\cite{haupt2023}
and a symmetric two-electron term $u_2({\bf r}_i,{\bf r}_j)$. For a fixed
$N$-electron system these combine into a single symmetric effective pair kernel,
\begin{equation}
\begin{aligned}
    u({\bf r}_i,{\bf r}_j)&=u_2({\bf r}_i,{\bf r}_j)+
    \frac{\chi({\bf r}_i)+\chi({\bf r}_j)}{N-1},\\
    \hat{\tau} &= \sum_i^N \chi({\bf r}_i)
    + \sum_{i<j}^{N} u_2({\bf r}_i,{\bf r}_j)
    = \sum_{i<j}^{N} u({\bf r}_i,{\bf r}_j)\\
    &= \tfrac{1}{2}\sum_{i\neq j}^{N} u({\bf r}_i,{\bf r}_j),
    \qquad F({\bf R}) = \prod_{i<j} e^{u({\bf r}_i,{\bf r}_j)},
\end{aligned}
    \label{eq:tau-def}
\end{equation}
where the second equality holds because each electron label appears in exactly
$N-1$ ordered pairs, so the $1/(N-1)$ factor in $u$ resums the one-body term
exactly. The one-body factor is thereby represented exactly within the same
pair-product form used by the integral kernels. The effective $u$ is real-valued and symmetric,
$u({\bf r}_i,{\bf r}_j)=u({\bf r}_j,{\bf r}_i)$, and collects the one-electron
nuclear-cusp correction together with the electron--electron, electron--nucleus,
and three-body electron--electron--nucleus contributions in $u_2$. In
the expansion below, $\mathcal{P}({\bf r}_{i_1},\dots,{\bf r}_{i_n})$ denotes the
symmetrizer over the electron coordinates listed as its arguments, i.e.\ the sum over
all $n!$ permutations of the $n$ electron labels.

Expanding $\bar{H}$ via the Baker--Campbell--Hausdorff formula, and using the fact
that $\hat{\tau}$ depends only on the electron positions and contains no gradient or
integral operators, the series terminates at the
second commutator and produces at most three-body operators:
\begin{widetext}
\begin{equation}
    \begin{aligned}
    \bar{H} &= \hat{H} + [\hat{H},\hat{\tau}] + \tfrac{1}{2}[[\hat{H},\hat{\tau}],\hat{\tau}]\\
        &= \hat{H}
            - \sum_{i<j} \mathcal{P}({\bf r}_i,{\bf r}_j)\!\left[
                \tfrac{1}{2}\nabla_i^{2} u({\bf r}_i,{\bf r}_j)
                + \tfrac{1}{2}\big(\nabla_i u({\bf r}_i,{\bf r}_j)\big)^{2}
                + \nabla_i u({\bf r}_i,{\bf r}_j)\cdot\nabla_i
            \right]\\
        & \quad - \sum_{i<j<k} \mathcal{P}({\bf r}_i,{\bf r}_j,{\bf r}_k)\,
          \nabla_i u({\bf r}_i,{\bf r}_j)\cdot\nabla_i u({\bf r}_i,{\bf r}_k).
    \end{aligned}
    \label{eq:Hbar-bch}
\end{equation}
\end{widetext}
Projected onto a set of one-electron orbitals $\{\phi_p\}$, $\bar{H}$ assumes the
second-quantized form
\begin{equation}
    \begin{aligned}
    \bar{H} =  {}&h_{pq}\, a_p^\dagger a_q
    + \tfrac{1}{2}\,(V^{pq}_{rs}-K^{pq}_{rs})\, a_p^\dagger a_q^\dagger a_s a_r\\
    &- \tfrac{1}{6}\,L^{pqr}_{stu}\, a_p^\dagger a_q^\dagger a_r^\dagger a_u a_t a_s ,
    \end{aligned}
    \label{eq:Hbar-2nd}
\end{equation}
where the Einstein summation convention is used. The two-body $K$ and three-body $L$
corrections are
\begin{widetext}
\begin{equation}
    \begin{aligned}
        K^{pq}_{rs} &= \hat{\mathcal{P}}_{12}\!\left[
        \Braket{\phi_p(1)\phi_q(2) |
        \tfrac{1}{2}\nabla_1^{2} u(1,2) + \tfrac{1}{2}(\nabla_1 u(1,2))^{2}
        + \nabla_1 u(1,2)\cdot\nabla_1 | \phi_r(1)\phi_s(2)}\right],\\
        L^{pqr}_{stu} &= \hat{\mathcal{P}}_{123}\,
        \Braket{\phi_p(1)\phi_q(2)\phi_r(3) |
            \nabla_1 u(1,2)\cdot\nabla_1 u(1,3)
            | \phi_s(1)\phi_t(2)\phi_u(3)},
    \end{aligned}
\end{equation}
\end{widetext}
with the shorthand $\phi_p(i) \equiv \phi_p({\bf r}_i)$ and $u(i,j)\equiv u({\bf r}_i,{\bf r}_j)$.

\subsection{Jastrow correlator used in this work}

We use the flexible multi-center Boys--Handy (BH) correlator\cite{boys1969},
which captures electron--electron, electron--nucleus, and three-body
electron--electron--nucleus correlation, with its linear and nonlinear
parameters optimized by variational Monte Carlo (VMC). The BH form is written not in
terms of the bare interparticle distances but in terms of bounded, monotone
\emph{scaled distances}
\begin{equation}
    \bar r_{iI} \;=\; \frac{b_I\,r_{iI}}{1 + b_I\,r_{iI}},
    \qquad
    \bar r_{ij} \;=\; \frac{d\,r_{ij}}{1 + d\,r_{ij}},
    \label{eq:bh-scaled}
\end{equation}
with ${\bf r}_{iI}={\bf r}_i-{\bf R}_I$, $r_{iI}=|{\bf r}_{iI}|$, $r_{ij}=|{\bf r}_i-{\bf r}_j|$, and variational
scale parameters $b_I>0$ (nucleus-resolved) and $d>0$. The scaled variables satisfy
$\bar r\in[0,1)$, so that monomials $\bar r^m$ remain bounded at long range, and all
positive integer powers can be mixed without divergence. The correlator reads
\begin{equation}
    u_2({\bf r}_i,{\bf r}_j) = \sum_{I}\sum_{k} c_{k}^{I}\,
    \Big[\bar r_{iI}^{m_k}\, \bar r_{jI}^{n_k} + \bar r_{jI}^{m_k}\, \bar r_{iI}^{n_k}\Big]\,
    \bar r_{ij}^{o_k},
    \label{eq:bh-u}
\end{equation}
where $(m_k,n_k,o_k)$ are user-chosen non-negative-integer exponents and $c_k^I$ are
the linear variational coefficients. A dedicated term with $m_k=n_k=0$ and
$o_k=1$ is retained in a cusp-normalized form, with its masked prefactor fixed so
that the summed short-range slope satisfies the opposite-spin electron--electron
Kato condition, $\partial u_2/\partial r_{ij}|_{r_{ij}\to0}=1/2$. The same
spin-independent two-electron correlator $u_2$ is applied to every electron pair, so it
satisfies the opposite-spin Kato cusp exactly but does not separately impose the
same-spin cusp condition ($\tfrac14$); all systems studied here are closed-shell. The remaining
$c_k^I$, together with the nonlinear scale parameters $d$ and $b_I$, are
optimized freely by VMC.
The one-electron term $\chi$ is the nuclear-cusp factor of
Ref.~\onlinecite{haupt2023}; the composite correlator used in production is
specified in the Computational Details. All correlator derivatives entering
the integral kernels are obtained by forward-mode automatic differentiation,
which is robust to arbitrary monomial degrees $(m_k,n_k,o_k)$.

\subsection{Variance optimization of the correlator by variational Monte Carlo}
\label{sec:vmc}

The linear coefficients $\{c_k^I\}$ and the nonlinear scale parameters
$\{b_I, d\}$ in Eqs.~(\ref{eq:bh-scaled})--(\ref{eq:bh-u}) are optimized by
variational Monte Carlo prior to the TC integral construction. Following the
reference-variance approach of Ref.~\onlinecite{haupt2023}, which has been shown to
optimize transcorrelated correlators more reliably than energy minimization, we
minimize the variance of
the local energy of the Slater--Jastrow trial wavefunction
$\Psi_T({\bf R}) = F({\bf R})\,\Phi_0({\bf R})$, with $\Phi_0$ the Hartree--Fock
reference determinant.

\paragraph{Sampling from the reference determinant.}
A distinctive feature of the \emph{reference-variance} protocol\cite{haupt2023} is that Monte Carlo
configurations are drawn from the reference density $|\Phi_0({\bf R})|^{2}$ rather than from the
Slater--Jastrow density $|\Psi_T({\bf R})|^{2}$. The sampling distribution is
therefore held fixed throughout the parameter update and only the
local energy
\begin{equation}
    E_{\rm loc}({\bf R};\bm\theta)
    \;=\; \frac{\hat H\,\Psi_T({\bf R})}{\Psi_T({\bf R})},
    \label{eq:Eloc}
\end{equation}
depends on the Jastrow parameters $\bm\theta$ through the trial wavefunction
$\Psi_T$, with the transcorrelated Hamiltonian $\bar H({\bm\theta})$ given by Eq.~\eqref{eq:Hbar-bch}. This fixed sampling distribution
has two operational advantages: the sampler need not be re-equilibrated after
each parameter update, and the walker state stores only Slater matrices and their
inverses. These are updated at cost $\mathcal{O}(N_{\rm el}^{2})$ per proposal
(see Supporting Information Section~\ref{si-sec:si-sampling}), while the Jastrow enters only through $E_{\rm loc}$ and never through
the Metropolis acceptance ratio.

\paragraph{Objective function.}
Concretely, at optimization step $s$ we sample $N_w$ walker positions
$\{{\bf R}_k^{(s)}\}_{k=1}^{N_w}$ from $|\Phi_0|^{2}$. Holding this batch fixed,
write $E_k(\bm\theta)=E_{\rm loc}({\bf R}_k^{(s)};\bm\theta)$ and define its raw
mean and mean absolute deviation (MAD) as
\begin{equation}
  \begin{aligned}
  m_s(\bm\theta)&=\frac{1}{N_w}\sum_k E_k(\bm\theta),\\
  d_s(\bm\theta)&=\frac{1}{N_w}\sum_k\left|E_k(\bm\theta)-m_s(\bm\theta)\right|.
  \end{aligned}
\end{equation}
For outlier protection, each local energy is clipped to the interval set by the raw
mean and MAD,
\begin{equation}
  \begin{split}
  \widetilde E_k(\bm\theta)=\operatorname{clip}\!\big(&E_k(\bm\theta),
  m_s(\bm\theta)-5d_s(\bm\theta),\\
  &m_s(\bm\theta)+5d_s(\bm\theta)\big).
  \end{split}
\end{equation}
The mean is then recomputed from the clipped energies, and the fixed-batch objective is
\begin{equation}
  \begin{aligned}
  \overline{\widetilde E}_s(\bm\theta)&=\frac{1}{N_w}\sum_k\widetilde E_k(\bm\theta),\\
  \sigma_s^2(\bm\theta)&=\frac{1}{N_w-1}\sum_k
  \left[\widetilde E_k(\bm\theta)-\overline{\widetilde E}_s(\bm\theta)\right]^2.
  \end{aligned}
  \label{eq:var-loss}
\end{equation}
No external reference energy enters the objective. The update in Eq.~\eqref{eq:gn}
evaluates the gradient of Eq.~\eqref{eq:var-loss} at $\bm\theta^{(s)}$. The variance
of Eq.~(\ref{eq:var-loss}) vanishes only if $\Phi_0$ is an exact (right)
eigenstate of $\bar H(\bm\theta)$; within the finite Boys--Handy parameterization the
optimizer instead seeks the lowest attainable variance, approaching this
zero-variance principle to the extent that the correlator can absorb the
transcorrelated Hamiltonian into the reference determinant.

\paragraph{Newton optimizer with Gauss--Newton curvature.}
Because the objective Eq.~(\ref{eq:var-loss}) has the structure of a sum of squared
residuals $r_k(\bm\theta)=\widetilde E_k(\bm\theta)-
\overline{\widetilde E}_s(\bm\theta)$, we use a
Newton-type update with Gauss--Newton curvature,
\begin{equation}
    \begin{aligned}
    C_{ij}[\bm\theta]
    &\;=\; \frac{1}{N_w}\sum_{k=1}^{N_w}
    \big(\partial_{\theta_i} r_k\big)\!\big(\partial_{\theta_j} r_k\big)
    \;+\; \lambda\,\delta_{ij},\\
    \Delta\bm\theta &\;=\; -\,\eta\,C^{-1}\nabla_{\bm\theta}\sigma_s^{2},
    \end{aligned}
    \label{eq:gn}
\end{equation}
with an adaptive Tikhonov shift $\lambda$ and an exact Cholesky-based solve of the
linear system. Gauss--Newton matches the quadratic structure of
Eq.~(\ref{eq:var-loss}) near the optimum and converges in far fewer steps than
plain stochastic gradient descent.

\paragraph{Optimization schedule.}
The two-stage warm-up/refinement optimization schedule and the Polyak--Ruppert
parameter averaging used to define the production correlator are described in
Supporting Information Section~\ref{si-sec:si-vmc-schedule}.

\subsection{Interpolative Separable Density-Fitting}
\label{sec:isdf}

The integration grid underlying the ISDF factorization is a Becke-partitioned atomic
grid\cite{becke1988} with Treutler--Ahlrichs radial nodes, Lebedev angular grids,
and Bragg--Slater radii, yielding points $\{{\bf r}_g\}_{g=1}^{N_g}$ and weights
$\{w_g\}$ that absorb the nuclear partitioning and radial Jacobian. We use PySCF
grid level~2 throughout, as in Ref.~\onlinecite{haupt2023}.

\paragraph{Low-rank structure.}
The approximation relies on a compact representation of orbital products. The
pointwise Coulomb kernel itself is not low rank on the grid, but the contracted
four-index tensor inherits low rank from the orbital overlap density
\begin{equation}
    \rho^{p}_{q}({\bf r}) \equiv \phi_p({\bf r})\,\phi_q({\bf r}),
\end{equation}
because $\mathrm{rank}(AB)\le \min[\mathrm{rank}(A),\mathrm{rank}(B)]$. The same
argument guarantees that the transcorrelated integrals $K$ and $L$ admit a low-rank
representation of rank at most $N_\mu$, where $N_\mu$ is the numerical rank of
$\rho^p_q({\bf r})$ on the chosen grid.

\paragraph{Pivot selection by pivoted Cholesky.}
We select the scalar-channel interpolation points
$\{{\bf r}^{(\phi)}_\mu\}_{\mu=1}^{N_\mu^{(\phi)}}$ by an iterative
\emph{pivoted Cholesky decomposition} (rather than by QR-column-pivoting\cite{lu2015,hu2017}
or $k$-means\cite{dong2018} as in previous ISDF implementations), in the spirit
of the pivoted-Cholesky grid-point selection introduced for least-squares tensor
hypercontraction\cite{matthews2020}. Fusing the orbital indices into a single
row index $pq$ and indexing the grid points by the column, the overlap-density matrix
$M_{(pq),g}=\phi_p({\bf r}_g)\phi_q({\bf r}_g)$ has the separable Gram
\begin{equation}
    \begin{aligned}
    (M^\top M)_{g g'} &= \Big(\sum_p \phi_p({\bf r}_g)\phi_p({\bf r}_{g'})\Big)^{2},\\
    D^{(\phi)}_{g} \equiv (M^\top M)_{gg} &= \Big(\sum_p \phi_p({\bf r}_g)^{2}\Big)^{2},
    \end{aligned}
\end{equation}
which avoids ever forming the $N_{\rm orb}^{2}\times N_g$ matrix $M$ explicitly.
In the actual quadrature implementation the pivot selection is performed on the
weight-scaled orbitals $\phi^w_p({\bf r}_g)=w_g^{1/2}\phi_p({\bf r}_g)$, so the
scalar Cholesky matrix is
\begin{equation}
    \begin{aligned}
    S^{(\phi)}_{gg'} &=
    \Big(\sum_p \phi^w_p({\bf r}_g)\phi^w_p({\bf r}_{g'})\Big)^2,\\
    D^{(\phi)}_g &=
    w_g^2\Big(\sum_p \phi_p({\bf r}_g)^2\Big)^2 .
    \end{aligned}
\end{equation}
Because the two-body $K^{(1)}$ kernel (in Section~\ref{sec:tc-integrals}), in which
the gradient acts on an orbital, involves the orbital--gradient product
$\phi_p({\bf r})\,\nabla\phi_q({\bf r})$, we perform a \emph{second} pivoted
Cholesky on the diagonal
\begin{equation}
    D^{(\nabla)}_g =
    w_g^2\Big(\sum_p \phi_p({\bf r}_g)^{2}\Big)\!
          \Big(\sum_{p,c} \big(\partial_c\phi_p({\bf r}_g)\big)^{2}\Big),
\end{equation}
with the corresponding column
\begin{equation}
    S^{(\nabla)}_{gg'} =
    \Big(\sum_p \phi^w_p({\bf r}_g)\phi^w_p({\bf r}_{g'})\Big)
    \Big(\sum_{p,c} \partial_c\phi^w_p({\bf r}_g)
                    \partial_c\phi^w_p({\bf r}_{g'})\Big).
\end{equation}
Both production decompositions run to prescribed ranks
$N_\mu^{(\phi)}$ and $N_\mu^{(\nabla)}$; there is no residual-tolerance stopping
criterion. The ranks are set from the chosen rank ratios and the two resulting
pivot sets are subsequently fused. The decomposition itself is matrix-free
(Section~\ref{sec:impl-numerics}).

\paragraph{Pivot fusion.}
We take the \emph{union} of the two pivot sets as the final interpolation set,
denote its cardinality by $N_\mu$, and relabel its points
$\{{\bf r}_\mu\}_{\mu=1}^{N_\mu}$. This
forces the scalar and gradient auxiliary bases to share a single pivot index $\mu$,
so that the pivot-point tensors
$C^p_q({\bf r}_\mu)=\phi_p({\bf r}_\mu)\phi_q({\bf r}_\mu)$ and
$\tilde{\bf C}^p_q({\bf r}_\mu)=\phi_p({\bf r}_\mu)\nabla\phi_q({\bf r}_\mu)$
appear simultaneously in the later contractions without any re-indexing or
interpolation between incompatible pivot sets. Because the gradient pivots are
largely distinct from the scalar ones, the fused set is roughly twice the size of
a single channel, but every intermediate in the $K$ and
$\Delta U$ assembly addresses a single auxiliary
dimension.

\paragraph{Auxiliary basis fitting.}
Given the fused pivots, ISDF approximates the densities as
\begin{equation}
    \begin{aligned}
    \rho^{p}_{q}({\bf r}) &\;\approx\; \sum_\mu \xi_\mu({\bf r})\,C^{p}_{q}({\bf r}_\mu),\\
    \phi_p({\bf r})\,\partial_c\phi_q({\bf r})
        &\;\approx\; \sum_\mu \tilde\xi^c_\mu({\bf r})\,
        \tilde C^{p,c}_{q}({\bf r}_\mu),\qquad c\in\{x,y,z\},
    \end{aligned}
    \label{eq:isdf-rho}
\end{equation}
where $\xi_\mu({\bf r})$ and the component-specific $\tilde\xi_\mu^{c}({\bf r})$
are determined by minimizing the Frobenius norm of the residual, following the
least-squares fitting strategy first introduced for tensor
hypercontraction\cite{parrish2012} and later adopted for
ISDF\cite{hu2017}. The
associated normal equations
\begin{equation}
    \sum_\nu (C C^\top)_{\mu\nu}\,\xi_\nu({\bf r})
    = \sum_{pq} C^p_q({\bf r}_\mu)\,\rho^p_q({\bf r}),
\end{equation}
are solved directly through a Cholesky factorization of the dense
$N_\mu\times N_\mu$ Gram matrix $(CC^\top)_{\mu\nu}=G_{\mu\nu}^{2}$, where
$G_{\mu\nu}=\sum_p\phi_p({\bf r}_\mu)\phi_p({\bf r}_\nu)$
(see Section~\ref{sec:impl-numerics} for the numerical safeguards). The gradient auxiliary basis $\tilde\xi_\mu^{c}$ is
obtained by solving the analogous system three times, one per Cartesian component,
but with the mixed normal-equation matrix
\begin{equation}
    (\tilde C^{c}\tilde C^{c\top})_{\mu\nu}
    =
    \Big(\sum_p \partial_c\phi_p({\bf r}_\mu)\partial_c\phi_p({\bf r}_\nu)\Big)
    \Big(\sum_q \phi_q({\bf r}_\mu)\phi_q({\bf r}_\nu)\Big),
\end{equation}
and a right-hand side formed from the corresponding gradient--scalar product at the
target grid point. Thus the scalar auxiliary uses scalar--scalar pivot factors,
whereas each Cartesian gradient auxiliary uses a gradient factor on one orbital
index and a scalar factor on the other.
Following traditional ISDF terminology we refer to $\xi$ and $\tilde\xi$ as the
\emph{auxiliary density} and \emph{auxiliary gradient density}, respectively.

\paragraph{Grid-batched streaming.}
The tensors $\xi_\mu({\bf r}_g)$ and $\tilde\xi_\mu({\bf r}_g)$ are the only objects
whose storage grows with the dense grid size $N_g$. They are therefore evaluated
in grid batches and, when necessary, streamed to disk rather than held in device
or host memory. During integral assembly, batches are brought back to the device
asynchronously by a depth-one prefetch mechanism, so that host-to-device transfer
overlaps with compute.

\begin{table}[t]
    \caption{Fixed-rank, weight-scaled pivoted-Cholesky selection (scalar branch).}
    \label{alg:pivot}
    \centering
    \begin{tabular}{@{}l@{}}
    \hline\hline
    \begin{minipage}{0.95\columnwidth}
    \vspace{3pt}
    \begin{algorithmic}[1]
    \State \textbf{Input:} $\phi_p({\bf r}_g)$, positive weights $w_g$, target rank $N_\mu^{(\phi)}$.
    \State $d_g \leftarrow w_g^2\big(\sum_p \phi_p({\bf r}_g)^{2}\big)^{2}$.
    \For{$\mu = 1, \ldots, N_\mu^{(\phi)}$}
        \State $g^{\star} \leftarrow \arg\max_g d_g$; record pivot ${\bf r}_\mu \leftarrow {\bf r}_{g^{\star}}$.
        \State $s_g \leftarrow w_gw_{g^\star}
        \big(\sum_p\phi_p({\bf r}_g)\phi_p({\bf r}_{g^\star})\big)^2$.
        \State $L_{g\mu}\leftarrow
        \big(s_g-\sum_{\lambda<\mu}L_{g\lambda}L_{g^\star\lambda}\big)/\sqrt{d_{g^\star}}$.
        \State $d_g \leftarrow \max(d_g-L_{g\mu}^{2},0)$.
    \EndFor
    \State \textbf{return} exactly $N_\mu^{(\phi)}$ pivots $\{{\bf r}_\mu\}$ and factor $L$.
    \end{algorithmic}
    \vspace{3pt}
    \end{minipage}\\
    \hline\hline
    \end{tabular}
\end{table}

\section{ISDF Representation of the TC Integrals}
\label{sec:tc-integrals}

\subsection{Two-body $K$ integrals}

The real-space expression for the two-body correction in
Ref.~\onlinecite{Cohen2019} separates naturally into three kernels:
\begin{equation}
    \begin{aligned}
        K^{pq(1)}_{rs} &= \Braket{\phi_p(1)\phi_q(2)| \nabla_1 u(1,2)\cdot\nabla_1 |\phi_r(1)\phi_s(2)},\\
        K^{pq(2)}_{rs} &= \Braket{\phi_p(1)\phi_q(2)| \nabla_1^{2} u(1,2) |\phi_r(1)\phi_s(2)},\\
        K^{pq(3)}_{rs} &= \Braket{\phi_p(1)\phi_q(2)| (\nabla_1 u(1,2))^{2} |\phi_r(1)\phi_s(2)}.
    \end{aligned}
\end{equation}
Collecting the three kernels with the symmetrization $\hat{\mathcal P}_{12}$, the two-body
kernel entering Eq.~(\ref{eq:Hbar-2nd}) is
\begin{equation}
    K^{pq}_{rs} = \hat{\mathcal P}_{12}\!\left[
    K^{pq(1)}_{rs} + \tfrac12 K^{pq(2)}_{rs} + \tfrac12 K^{pq(3)}_{rs}\right],
    \label{eq:K-recombine}
\end{equation}
with $\hat{\mathcal P}_{12}\,T^{pq}_{rs}=T^{pq}_{rs}+T^{qp}_{sr}$ denoting
electron-1$\leftrightarrow$2 exchange, which simultaneously interchanges
$p\leftrightarrow q$ and $r\leftrightarrow s$ in the physicists' index convention used here.
To avoid the numerically expensive Laplacian, we integrate by parts, a
reformulation first used in the supplementary material of
Ref.~\onlinecite{Cohen2019}, to rewrite
\begin{equation}
    \begin{aligned}
        K^{pq(2)}_{rs} &= -\,\Braket{\phi_p(1)\phi_q(2)|\nabla_1 u(1,2)\cdot\nabla_1|\phi_r(1)\phi_s(2)}\\
        &\quad -\,\Braket{\phi_r(1)\phi_s(2)|\nabla_1 u(1,2)\cdot\nabla_1|\phi_p(1)\phi_q(2)},
    \end{aligned}
\end{equation}
which reduces $K^{(2)}$ to two $K^{(1)}$-style evaluations with swapped orbital labels; the
surface terms vanish for the localized square-integrable orbital products used here.
Writing $K^{(1,\mathrm{swap})pq}_{rs}=\Braket{\phi_r(1)\phi_s(2)|\nabla_1 u(1,2)\cdot\nabla_1|\phi_p(1)\phi_q(2)}$
for the swapped contraction, so that $K^{(2)}=-K^{(1)}-K^{(1,\mathrm{swap})}$, Eq.~(\ref{eq:K-recombine})
reduces to the form assembled in practice,
\begin{equation}
    K^{pq}_{rs} = \tfrac12\,\hat{\mathcal P}_{12}\!\left[
    K^{pq(1)}_{rs} - K^{pq(1,\mathrm{swap})}_{rs} + K^{pq(3)}_{rs}\right].
    \label{eq:K-recombine-final}
\end{equation}
We therefore discuss only $K^{(1)}$ and $K^{(3)}$ in what follows.

Substituting the ISDF factorizations in Eq.~(\ref{eq:isdf-rho}) yields
\begin{equation}
    \begin{aligned}
        K^{pq(1)}_{rs}
        &= \sum_{g,g'} w_g w_{g'}\,
           \big[\phi_p({\bf r}_g)\,\nabla\phi_r({\bf r}_g)\big]\!\cdot\!\nabla u({\bf r}_g,{\bf r}_{g'})\\
        &\qquad\times \phi_q({\bf r}_{g'})\phi_s({\bf r}_{g'})\\
        &= \sum_{\mu\nu} {\bf u}^{(1)}_{\mu\nu}\!\cdot\!
           \tilde{\bf C}^{p}_{r}({\bf r}_\mu)\,C^{q}_{s}({\bf r}_\nu),
    \end{aligned}
    \label{eq:K1-isdf}
\end{equation}
with the \emph{vector-valued} Jastrow kernel (three Cartesian components)
\begin{equation}
    \begin{aligned}
    u^{(1),c}_{\mu\nu}
    &= \sum_{g,g'} w_g\,w_{g'}\,\tilde\xi^c_\mu({\bf r}_g)\,
       \xi_\nu({\bf r}_{g'})\,\partial_c u({\bf r}_g,{\bf r}_{g'}),\\
    {\bf u}^{(1)}&\in\mathbb{R}^{N_\mu\times N_\mu\times 3}.
    \end{aligned}
\end{equation}
The analogous $K^{(3)}$ expression reads
\begin{equation}
    \begin{aligned}
    K^{pq(3)}_{rs} &= \sum_{\mu\nu} u^{(3)}_{\mu\nu}\,
    C^{p}_{r}({\bf r}_\mu)\,C^{q}_{s}({\bf r}_\nu),\\
    u^{(3)}_{\mu\nu} &= \sum_{g,g'} w_g w_{g'}\,\xi_\mu({\bf r}_g)\,\xi_\nu({\bf r}_{g'})\,
    \big(\nabla u({\bf r}_g,{\bf r}_{g'})\big)^{2},
    \end{aligned}
    \label{eq:K3-isdf}
\end{equation}
where now $u^{(3)}\in\mathbb{R}^{N_\mu\times N_\mu}$ is \emph{scalar} (no Cartesian
component) because it contracts the squared gradient.

\subsection{Effective two-body correction $\Delta U$ from xTC}

Explicit evaluation of the rank-six $L$ tensor in Eq.~(\ref{eq:Hbar-2nd}) is
prohibitive already for moderate system sizes. Following the xTC scheme of
Ref.~\onlinecite{christlmaier2023} we instead absorb $L$ into an effective two-body
correction by contracting two of its three pairs with the Hartree--Fock one-particle reduced density matrix
$\gamma^{u}_{t}=\Braket{a_t^\dagger a_u}$, defined here as the one-spin
spatial-orbital density (occupied eigenvalues equal to one). We keep the
physicists' index convention used throughout the article: in
$\Delta U^{pq}_{rs}$, the pairs $(p,r)$ and $(q,s)$ are the bra--ket orbital
pairs of electron 1 and electron 2, respectively. The result is a sum of four contributions
that, after inserting the ISDF factorizations~(\ref{eq:isdf-rho}), each take a form
well suited to grid-batched contraction:
\begin{equation}
    \Delta U^{pq}_{rs}
    = -\,\hat{\mathcal P}_{12}\!\big[\,
        \mathrm{Term}_1 + \mathrm{Term}_2 + \mathrm{Term}_3 + \mathrm{Term}_4\,\big],
    \label{eq:deltaU-four}
\end{equation}
with the electron-1$\leftrightarrow$2 exchange
$\hat{\mathcal P}_{12}T^{pq}_{rs}=T^{pq}_{rs}+T^{qp}_{sr}$ acting as in
Eq.~(\ref{eq:K-recombine}), which enforces the particle-exchange relation
$\Delta U^{pq}_{rs}=\Delta U^{qp}_{sr}$. To specify the four terms before applying ISDF, define
the orbital-pair and density-matrix-contracted fields
\begin{equation}
\begin{aligned}
 \rho^a_b({\bf r}_g)&=\phi_a({\bf r}_g)\phi_b({\bf r}_g),\\
 \rho_\gamma({\bf r}_g)&=\sum_{tu}\gamma^u_t\rho^t_u({\bf r}_g),\\
 \beta^u_s({\bf r}_g)&=\sum_t\gamma^u_t\rho^t_s({\bf r}_g),\\
 {\bf A}^{a}_{b,g}&=\sum_{g'}w_{g'}\rho^a_b({\bf r}_{g'})
       \nabla u({\bf r}_g,{\bf r}_{g'}),\\
 {\bf A}_{\gamma,g}&=\sum_{g'}w_{g'}\rho_\gamma({\bf r}_{g'})
       \nabla u({\bf r}_g,{\bf r}_{g'}),\\
 {\bf B}^{u}_{s,g}&=\sum_{g'}w_{g'}\beta^u_s({\bf r}_{g'})
       \nabla u({\bf r}_g,{\bf r}_{g'}).
\end{aligned}
\label{eq:deltaU-fields}
\end{equation}
The pre-ISDF contributions are then
\begin{equation}
\begin{aligned}
 \mathrm{Term}_1 &=2\sum_g w_g\rho^p_r({\bf r}_g)
       {\bf A}_{\gamma,g}\!\cdot\!{\bf A}^{q}_{s,g},\\
 \mathrm{Term}_2 &=-\sum_g w_g\rho^p_r({\bf r}_g)
       \sum_u {\bf B}^{u}_{s,g}\!\cdot\!{\bf A}^{q}_{u,g},\\
 \mathrm{Term}_3 &=-\sum_g w_g{\bf A}^{p}_{r,g}\!\cdot\!
       \sum_u\!\left[\beta^u_s({\bf r}_g){\bf A}^{q}_{u,g}
       +{\bf B}^{u}_{s,g}\rho^q_u({\bf r}_g)\right],\\
 \mathrm{Term}_4 &=\sum_g w_g\rho_\gamma({\bf r}_g)
       {\bf A}^{p}_{r,g}\!\cdot\!{\bf A}^{q}_{s,g}.
\end{aligned}
\label{eq:deltaU-preisdf}
\end{equation}
Throughout this section, $\mu,\nu,\lambda$ label auxiliary (ISDF) indices, so that
${\bf r}_\mu$ denotes the corresponding pivot point; $g,g'$ label grid points,
$c\in\{x,y,z\}$ labels Cartesian components, $\alpha$ labels columns of the
density projector, and $p,q,r,s,t,u$ label MO indices. We keep the pivot-tensor notation of Eq.~(\ref{eq:isdf-rho}), writing
scalar factors as $C^{p}_{q}({\bf r}_\mu)$ and reserving
$\tilde{\bf C}^{p}_{q}({\bf r}_\mu)$ for gradient factors. We pre-compute three
orbital-independent intermediates,
\begin{equation}
    \begin{aligned}
    G_{\nu} &= \sum_{tu} C^{t}_{u}({\bf r}_\nu)\,\gamma^{u}_{t},\\
    L^{u}_{s\nu} &= \sum_t \gamma^{u}_{t}\,C^{t}_{s}({\bf r}_\nu),\\
    \mathcal{G}^{c}_{\mu,g} &= \sum_{g'} w_{g'}\,\xi_\mu({\bf r}_{g'})\,\partial_c u({\bf r}_g,{\bf r}_{g'}).
    \end{aligned}
\end{equation}
Here $G_\nu$ is the density matrix contracted against an ISDF vertex,
$L^{u}_{s\nu}$ is the corresponding density-dressed vertex with one open orbital
pair, and $\mathcal G^c_{\mu,g}$ is a grid-local
\emph{gradient--interaction kernel} encoding the action of $\nabla u$ on the
auxiliary basis. Inserting Eq.~(\ref{eq:isdf-rho}) into
Eq.~(\ref{eq:deltaU-fields}) gives the direct mapping
\begin{equation}
\begin{aligned}
 \rho_\gamma({\bf r}_g)&\simeq\sum_\lambda G_\lambda\xi_\lambda({\bf r}_g),
 &{\bf A}^{a}_{b,g}&\simeq\sum_\mu C^a_b({\bf r}_\mu)
          {\boldsymbol{\mathcal G}}_{\mu,g},\\
 \beta^u_s({\bf r}_g)&\simeq\sum_\nu L^{u}_{s\nu}\xi_\nu({\bf r}_g),
 &{\bf B}^{u}_{s,g}&\simeq\sum_\nu L^{u}_{s\nu}
          {\boldsymbol{\mathcal G}}_{\nu,g}.
\end{aligned}
\label{eq:deltaU-isdf-map}
\end{equation}
Thus Terms~1 and 4 generate the $D^{(1)}$ and $D^{(4)}$ kernels below, while
Terms~2 and 3 generate $X^{(2)}$ and $X^{(3)}$, respectively.

\paragraph{Terms 1 and 4 (factorizable).}
Both terms reduce to a pure $N_\mu^{2}$ contraction, with dominant grid cost
$\mathcal{O}(N_g N_\mu^{2})$:
\begin{equation}
    \begin{aligned}
        \mathrm{Term}_4
        &= \sum_{\mu\nu} C^{p}_{r}({\bf r}_\mu)\,D^{(4)}_{\mu\nu}\,
           C^{q}_{s}({\bf r}_\nu),\\
        D^{(4)}_{\mu\nu}
        &= \sum_g \tilde{w}_g
           \!\left(\sum_c \mathcal{G}^{c}_{\mu,g}\mathcal{G}^{c}_{\nu,g}\right),\\
        \tilde{w}_g &= w_g\sum_\lambda G_\lambda\,\xi_\lambda({\bf r}_g),\\
        \mathrm{Term}_1
        &= 2\sum_{\mu\nu} C^{p}_{r}({\bf r}_\mu)\,D^{(1)}_{\mu\nu}\,
           C^{q}_{s}({\bf r}_\nu),\\
        D^{(1)}_{\mu\nu}
        &= \sum_g w_g\,\xi_\mu({\bf r}_g)\,V_{\nu,g},\\
        V_{\nu,g} &= \sum_c H_{g c}\,\mathcal{G}^{c}_{\nu,g},
           \quad H_{gc}=\sum_\lambda G_\lambda\mathcal{G}^{c}_{\lambda,g}.
    \end{aligned}
\end{equation}
The intermediates $D^{(4)}$ and $D^{(1)}$ are of $\mathcal{O}(N_\mu^{2})$.

\paragraph{Terms 2 and 3 (non-factorizable).}
These contributions retain one open orbital index inherited from the density matrix.
Introducing
\begin{equation}
    \begin{aligned}
    Y^{u}_{s,g,c} &= \sum_\nu L^{u}_{s\nu}\,\mathcal{G}^{c}_{\nu,g},\\
    Z^{q}_{u,g,c} &= \sum_\nu C^{q}_{u}({\bf r}_\nu)\,\mathcal{G}^{c}_{\nu,g},
    \end{aligned}
\end{equation}
one obtains
\begin{equation}
    \begin{aligned}
        \mathrm{Term}_2 &= -\sum_\mu C^{p}_{r}({\bf r}_\mu)\,X^{(2)}_{qs\mu},\\
        X^{(2)}_{qs\mu} &= \sum_g w_g\,\xi_\mu({\bf r}_g)\sum_{u,c} Y^{u}_{s,g,c}\,Z^{q}_{u,g,c},\\
        \mathrm{Term}_3 &= -\sum_\mu C^{p}_{r}({\bf r}_\mu)\,X^{(3)}_{qs\mu},
        \quad X^{(3)}_{qs\mu} = X^{(3,1)}_{qs\mu} + X^{(3,2)}_{qs\mu},
    \end{aligned}
\end{equation}
with
\begin{equation}
    \begin{aligned}
        X^{(3,1)}_{qs\mu} &= \sum_{g,c} w_g\,\mathcal{G}^{c}_{\mu,g}\sum_u \tilde L^{u}_{s,g}\,Z^{q}_{u,g,c},\\
        \tilde L^{u}_{s,g} &= \sum_\nu L^{u}_{s\nu}\,\xi_\nu({\bf r}_g),\\
        X^{(3,2)}_{qs\mu} &= \sum_{g,c} w_g\,\mathcal{G}^{c}_{\mu,g}\sum_u Y^{u}_{s,g,c}\,\widehat C^{q}_{u,g},\\
        \widehat C^{q}_{u,g} &= \sum_\nu C^{q}_{u}({\bf r}_\nu)\,\xi_\nu({\bf r}_g).
    \end{aligned}
\end{equation}
The dominant grid cost of Terms~2--3 is $\mathcal{O}(N_g N_\mu N_{\rm orb}^{2})$,
set by the contraction over the open orbital index $u$; the storage of the
resulting intermediates and their disk staging are discussed in
Section~\ref{sec:pipeline}.

\paragraph{Low-rank projector for Terms 2 and 3.}
A key algorithmic ingredient of our implementation is that the density-matrix
contraction inside the exchange kernel can be applied through a low-rank projector.
Observing that
\begin{equation}
    \begin{aligned}
    \sum_{u,c} Y^{u}_{s,g,c}\,Z^{q}_{u,g,c}
    &= \sum_{u,c}\sum_{\mu,\nu}L^{u}_{s\nu}\,\mathcal G^{c}_{\nu,g}\,C^{q}_{u}({\bf r}_\mu)\,\mathcal G^{c}_{\mu,g}\\
    &= \sum_{\mu\nu}\Big[\sum_u L^{u}_{s\nu} C^{q}_{u}({\bf r}_\mu)\Big]\!\Big(\sum_c \mathcal G^{c}_{\nu,g}\mathcal G^{c}_{\mu,g}\Big),
    \end{aligned}
\end{equation}
the bracketed density contraction can be written in factored form by decomposing
the reference density matrix as $\gamma = L_Q L_Q^\top$, with $L_Q$ the code-level
low-rank projector:
\begin{equation}
    \begin{aligned}
    \sum_u L^{u}_{s\nu} C^{q}_{u}({\bf r}_\mu)
      &= \sum_{\alpha} P_{s\nu,\alpha}P_{q\mu,\alpha},\\
    P_{s\nu,\alpha} &= \sum_t C^{t}_{s}({\bf r}_\nu)(L_Q)_{t\alpha},\\
    P_{q\mu,\alpha} &= \sum_u C^{q}_{u}({\bf r}_\mu)(L_Q)_{u\alpha}.
    \end{aligned}
\end{equation}
For the closed-shell Hartree--Fock reference used here, $\gamma$ is idempotent in
the MO basis, so $L_Q$ coincides with the density projector itself. The naive
$\mathcal O(N_g N_\mu^{2})$ exchange contraction is thereby replaced by an
$\mathcal O(N_g N_\mu N_{\rm orb})$ contraction; the measured speedup is
reported in Section~\ref{sec:impl-numerics}.

\paragraph{Final assembly.}
Combining the four terms,
\begin{equation}
    D_{\mu\nu} \equiv D^{(4)}_{\mu\nu}+2D^{(1)}_{\mu\nu},\qquad
    X_{qs\mu} \equiv X^{(2)}_{qs\mu}+X^{(3)}_{qs\mu},
    \label{eq:delta-u-dx-kernels}
\end{equation}
the final assembly is
\begin{equation}
\begin{aligned}
    \Delta U^{pq}_{rs}
    = -\hat{\mathcal P}_{12}\!\Big[&
        \sum_{\mu\nu}C^{p}_{r}({\bf r}_\mu)D_{\mu\nu}C^{q}_{s}({\bf r}_\nu)\\
        &- \sum_\mu C^{p}_{r}({\bf r}_\mu)X_{qs\mu}
    \Big].
\end{aligned}
\end{equation}
The assembly
scales as $\mathcal{O}(N_{\rm orb}^{4}N_\mu)$ in work and produces
$\mathcal{O}(N_{\rm orb}^{4})$ CCSD-facing tensor blocks when those blocks are
materialized---identical in logical size to the final contraction step of a
conventional density-fitted ERI assembly---and dominates the end-to-end cost for
large $N_{\rm orb}$.

\begin{table}[t]
\caption{ISDF-xTC: assembly of the effective transcorrelated Hamiltonian.}
\label{alg:isdfxtc}
\centering
\begin{tabular}{@{}l@{}}
\hline\hline
\begin{minipage}{0.95\columnwidth}
\vspace{3pt}
\begin{algorithmic}[1]
    \State Obtain molecular orbitals $\{\phi_p({\bf r})\}$ (Hartree--Fock or MP2 natural orbitals) and reference density matrix $\gamma^{u}_{t}$.
    \State Build Becke atomic grid $\{{\bf r}_g, w_g\}$; evaluate $\phi_p$ and $\nabla\phi_p$ on the grid.
    \State \textbf{Pivot selection:} weighted pivoted Cholesky on $\phi\phi$ and $\phi\nabla\phi$; fuse pivot sets.
    \State \textbf{Auxiliary fit:} solve normal equations for $\xi_\mu({\bf r})$ and $\tilde\xi_\mu^c({\bf r})$ (batched Cholesky with adaptive jitter, optional HDF5 streaming).
    \State Form pivot-point tensors $C^{p}_{q}({\bf r}_\mu)$, $\tilde{\bf C}^{p}_{q}({\bf r}_\mu)$.
    \State \textbf{$K$ kernels:} assemble ${\bf u}^{(1)}_{\mu\nu}$, $u^{(3)}_{\mu\nu}$ with doubly-blocked grid sums; contract into $K^{(1)},K^{(3)}$.
    \State \textbf{xTC correction:} form the derivation-defined $G_\nu$,
    $\mathcal G^c_{\mu,g}$, $L^{u}_{s\nu}$, $Y^{u}_{s,g,c}$, and $Z^{q}_{u,g,c}$;
    assemble $D^{(1)}$, $D^{(4)}$, $X^{(2)}$, and $X^{(3)}$ (including
    $\tilde L^{u}_{s,g}$ and $\widehat C^{q}_{u,g}$) on streamed grid shards using the
    low-rank projector and upper-triangular symmetry.
    \State Assemble $\Delta U^{pq}_{rs}$.
    \State Pass the transformed one- and two-body integrals $h+\Delta h$, $V-K+\Delta U$ to the CCSD solver.
\end{algorithmic}
\vspace{3pt}
\end{minipage}\\
\hline\hline
\end{tabular}
\end{table}

\subsection{Assembly of the effective xTC Hamiltonian}
\label{sec:xtc-ham}

The two-body kernels of the preceding sections deliver the corrections
$K^{pq}_{rs}$ and $\Delta U^{pq}_{rs}$. Contracting these against the
closed-shell reference one-particle density matrix reduces the transcorrelated
Hamiltonian to a standard zero-, one-, and two-body operator, which is the
object handed to the coupled-cluster solver. It is convenient to collect the
two-body correction into
\begin{equation}
    W^{pq}_{rs} \equiv -\,K^{pq}_{rs} + \Delta U^{pq}_{rs},
\end{equation}
so that the effective two-body integrals are $V^{pq}_{rs} + W^{pq}_{rs}$.

\paragraph{One-body correction and Fock operator.}
A partial contraction of $\Delta U$ over the occupied orbitals defines the
one-body increment
\begin{equation}
    \Delta h_{pq}
    = -\tfrac{1}{2}\sum_{o\in\mathrm{occ}}
      \big(2\,\Delta U^{po}_{qo} - \Delta U^{po}_{oq}\big).
    \label{eq:xtc-deltah}
\end{equation}
The effective one-body (Fock) operator adds to $\Delta h$ the mean-field
contraction of the two-body correction $W$ over the occupied space,
\begin{equation}
    f^{\rm xTC}_{pq}
    = \Delta h_{pq}
      + \sum_{i\in\mathrm{occ}}
        \big(2\,W^{pi}_{qi} - W^{pi}_{iq}\big),
    \label{eq:xtc-fock}
\end{equation}
comprising a Coulomb-type term (weight $2$ for the closed shell) and an
exchange-type term. This correction is added to the bare-Hamiltonian Fock
matrix built from $V$; because $K$ and $\Delta U$ are non-Hermitian,
$f^{\rm xTC}$ is stored without symmetrization.

\paragraph{Scalar constant.}
The fully-contracted remainder is a scalar that shifts the reference energy,
\begin{equation}
    E^{\rm xTC}_{0}
    = -\tfrac{2}{3}\sum_{i\in\mathrm{occ}}\Delta h_{ii} + E_{\rm nuc},
    \label{eq:xtc-const}
\end{equation}
formed from the same one-body operator $\Delta h$ traced over the occupied
orbitals, together with the nuclear repulsion $E_{\rm nuc}$.

\paragraph{Assembled Hamiltonian.}
The Hamiltonian handed to the solver retains the original one- and two-electron
integrals, with the one-body part renormalized by $\Delta h$ and the two-body
part xTC-corrected,
\begin{widetext}
\begin{equation}
    \bar H_{\rm xTC}
    = E^{\rm xTC}_0
      + \big(h_{pq} + \Delta h_{pq}\big)\, a_p^\dagger a_q
      + \tfrac{1}{2}\big(V^{pq}_{rs} - K^{pq}_{rs} + \Delta U^{pq}_{rs}\big)\,
        a_p^\dagger a_q^\dagger a_s a_r ,
    \label{eq:xtc-hamiltonian}
\end{equation}
\end{widetext}
where $h_{pq}$ are the bare one-electron integrals and $\Delta h$ and
$E^{\rm xTC}_0$ are given by Eqs.~\eqref{eq:xtc-deltah}
and~\eqref{eq:xtc-const}. From this $\bar H_{\rm xTC}$ the coupled-cluster solver
builds its reference Fock matrix in the usual closed-shell way; the xTC
contribution to that Fock is $f^{\rm xTC}$ of Eq.~\eqref{eq:xtc-fock}. Because
$K$ and $\Delta U$ are non-Hermitian, all one- and two-body integrals are stored
without symmetrization.

\section{Implementation}

\subsection{Integration with CCSD}

The transformed one- and two-body integrals,
$\tilde h_{pq}=h_{pq}+\Delta h_{pq}$ and
$\tilde V^{pq}_{rs}=V^{pq}_{rs}-K^{pq}_{rs}+\Delta U^{pq}_{rs}$,
respectively, are supplied to a restricted CCSD solver adapted to the
non-Hermitian transcorrelated Hamiltonian. The transformed two-body correction
is $-K+\Delta U$. It preserves
particle exchange, $W^{pq}_{rs}=W^{qp}_{sr}$, but breaks the Hermitian
pair-block relation $W^{pq}_{rs}=W^{rs}_{pq}$ and hence the bare ERI's eight-fold
symmetry. The solver therefore treats non-equivalent integral blocks as independent tensors
rather than relating them by transposition. The one-body Fock
matrix and the scalar constant are assembled as described in
Section~\ref{sec:xtc-ham}; like the two-body blocks they are non-Hermitian and
stored without symmetrization.

Although carrying the ISDF factorization directly through the CCSD amplitude
equations can yield lower-scaling formulations, that reduction is not used in the
present study. Here, ISDF is used to construct the transformed integral blocks,
which are then supplied to a conventional projective CCSD contraction sequence,
modified only as required to retain the independent non-Hermitian blocks described
above. The factor-direct formulation now under development retains the full
doubles-amplitude tensor but contracts its virtual indices sequentially with the
ISDF factors, applies the central kernels in the resulting
occupied-pair/interpolation-rank intermediates, and back-transforms sequentially.
This avoids explicit construction of the four-virtual integral tensor and reduces
the dominant contraction from nominal sixth- to fifth-order scaling, without
introducing a separate factorization of the doubles amplitudes. Development and
validation of this lower-scaling solver will be pursued in future work.

The complete non-equivalent block list, together with the block-by-block tiling,
streaming, and on-the-fly $(vvvv)$ protocol, is given in Supporting Information
Section~\ref{si-sec:si-ccsd-streaming}.

\subsection{Numerical implementation notes}
\label{sec:impl-numerics}

\paragraph{Correlator evaluation.}
The one-electron term $\chi$ is evaluated through the same ordered-pair
interface as $u_2$, entering each ordered pair as $\chi({\bf r}_i)/(N-1)$;
summing its derivative over $j\ne i$ reconstructs $\nabla_i\chi({\bf r}_i)$
exactly, so this evaluation is equivalent to the symmetric effective kernel of
Eq.~(\ref{eq:tau-def}).

\paragraph{VMC numerics.}
The fixed reference density $|\Phi_0|^{2}$ is sampled with Gaussian
single-electron Metropolis--Hastings proposals and Sherman--Morrison
determinant updates. The parameter-space dimension is modest for the BH
correlator (typically $N_{\rm p}\lesssim 100$), so the dense Gauss--Newton
curvature matrix of Eq.~(\ref{eq:gn}) is stored and factorized directly.

\paragraph{Matrix-free pivot search.}
The pivoted Cholesky selection of Section~\ref{sec:isdf} never materializes the
orbital-pair matrix $M$ or the dense grid-by-grid Gram matrix $S$. After a
pivot $g^\star$ is chosen, the new Cholesky column is
$L_{g\mu}=(S_{g g^\star}-\sum_{\lambda<\mu}L_{g\lambda}
L_{g^\star\lambda})/\sqrt{d_{g^\star}}$, and the residual diagonal is updated
as $d_g\leftarrow d_g-L_{g\mu}^2$. Each iteration stores only the residual
diagonal, the previously accepted Cholesky columns, and one newly generated
column $S_{g g^\star}$ computed from orbital values (and, for the gradient
branch, orbital gradients) at the candidate pivot and all grid points. This
keeps the pivot search linear in $N_g$ per accepted pivot and avoids the
$N_{\rm orb}^2 N_g$ storage that an explicit orbital-pair matrix would
require.

\paragraph{Auxiliary-fit robustness.}
Should the Cholesky factorization of the fitting Gram matrix of
Section~\ref{sec:isdf} fail due to near-singularity, a small adaptive Tikhonov
shift is added to the diagonal and the decomposition is retried.

\paragraph{Measured impact of the low-rank projector.}
In our benchmarks the low-rank projector of Section~\ref{sec:tc-integrals}
yields roughly a five-fold end-to-end wall-clock speedup (up to nine-fold for
the isolated $Q$-matrix contraction) in a benchmark using
$N_{\rm ch}\approx 10\,N_{\rm orb}$ per channel
($N_\mu\sim 20\,N_{\rm orb}$ after fusion), while preserving agreement with
the dense reference tensor to $\mathcal O(10^{-10})$ relative Frobenius
error.

\subsection{Multi-GPU pipeline and out-of-core storage}
\label{sec:pipeline}

\paragraph{Scaling and memory management.}
The kernels ${\bf u}^{(1)}$ and $u^{(3)}$ are orbital-independent and assembled once.
Their construction scales as $\mathcal{O}(N_g^{2} N_\mu)$ (using each auxiliary
vector only once per grid pair) and storage is $\mathcal{O}(N_\mu^{2})$. The final
assembly of the full four-index tensor in Eqs.~(\ref{eq:K1-isdf})--(\ref{eq:K3-isdf})
first contracts one auxiliary index at
$\mathcal{O}(N_\mu^{2}N_{\rm orb}^{2})$ cost and then contracts the remaining
auxiliary index into the four open orbital indices at
$\mathcal{O}(N_\mu N_{\rm orb}^{4})$ cost. The latter is the dominant step whenever
four-index CCSD-facing blocks are materialized. Those blocks have
$\mathcal{O}(N_{\rm orb}^{4})$ logical size, but production assembly streams panels
of width $B$ in one orbital index, giving peak panel storage
$\mathcal{O}(B N_{\rm orb}^{3})$ in addition to the
$\mathcal{O}(N_\mu^{2})$ kernels rather than holding the full tensor in memory.
By contrast, a direct dense grid evaluation of the same integrals without density
fitting would scale as $\mathcal{O}(N_g^{2} N_{\rm orb}^{2})$, carrying the full
grid-pair sum into every orbital-pair contraction; the ISDF compression
$N_{\rm orb}^{2}\!\to\!N_\mu$ (with $N_\mu$ a small multiple of $N_{\rm orb}$) and
the removal of the dense grid index from the final tensors are what render the
construction tractable.
Thus the low-rank compression removes the dense grid dependence from the
transcorrelated-kernel intermediates, while the final solver interface is still
handled as tiled or streamed four-index data. To keep the peak working memory
independent of $N_g$, we organize both grid sums as doubly-blocked loops over the bra
and ket grid batches, and block the pivot-to-orbital contraction over the auxiliary
index $\mu$; Jastrow derivatives are obtained by forward-mode automatic
differentiation of the correlator at every grid-pair batch.

\begin{table*}[t]
    \caption{Dominant cost and storage of the ISDF-xTC pipeline. Storage entries
    denote peak intermediate storage except where final four-index solver blocks
    are materialized or staged as logical CCSD-facing tensors.}
    \label{tab:scaling}
    \centering
    \begin{tabular}{lll}
        \hline\hline
        Step & Dominant work & Peak/logical storage \\ \hline
        Pivoted Cholesky         & $\mathcal{O}(N_g N_\mu^{2})$            & $\mathcal{O}(N_g N_\mu)$ \\
        Auxiliary fit (batched)  & $\mathcal{O}(N_g N_\mu^{2})$            & $\mathcal{O}(N_\mu^{2})$ \\
        $K^{(1)}, K^{(3)}$ kernels & $\mathcal{O}(N_g^{2} N_\mu)$          & $\mathcal{O}(N_\mu^{2})$ \\
        $K$ four-index assembly  & $\mathcal{O}(N_\mu^{2}N_{\rm orb}^{2}+N_\mu N_{\rm orb}^{4})$ & $\mathcal{O}(N_\mu^2+B N_{\rm orb}^{3})$ peak; $\mathcal{O}(N_{\rm orb}^{4})$ logical \\
        $\Delta U$ Terms 1,4     & $\mathcal{O}(N_g N_\mu^{2})$            & $\mathcal{O}(N_\mu^{2})$ \\
        $\Delta U$ Terms 2,3 (with low-rank projector) & $\mathcal{O}(N_g N_\mu N_{\rm orb})$ & $\mathcal{O}(N_{\rm orb}^{2} N_\mu)$ \\
        $\Delta U$ assembly      & $\mathcal{O}(N_{\rm orb}^{4} N_\mu)$    & $\mathcal{O}(N_{\rm orb}^{4})$ \\
        \hline\hline
    \end{tabular}
\end{table*}

The asymptotic scaling in Table~\ref{tab:scaling} hides a practical bottleneck:
several blocks of $\Delta U$ are too large to hold in GPU memory.
For an illustrative all-electron benzene-scale sizing case ($N_{\rm occ}=21$, $N_{\rm vir}\approx 200$,
double precision) the $(ovvv)$ and $(vovv)$ panels of $\Delta U$ are each about
$1.3$~GB and the dense $(vvvv)$ panel is about $13$~GB; by
$N_{\rm vir}\approx 300$, these grow to about $4.5$~GB and $65$~GB, respectively.
The $(ovvv)$ and $(vovv)$ panels are therefore staged to disk and, during CCSD,
streamed back to device memory one orbital tile at a time, while the dense
$(vvvv)$ panel is either contracted on the fly or, when a materialized $(vvvv)$ is
required, staged to disk through the same pipeline. The exchange-kernel
intermediate $X^{(2)}_{qs\mu}$ of Section~\ref{sec:tc-integrals} is disk-staged
the same way: it carries $\mathcal{O}(N_{\rm orb}^{2}N_\mu)$ storage, which for
benzene at cc-pCV5Z ($N_{\rm orb}=1200$, $N_\mu\approx 26{,}000$) would occupy
$\sim 300$~GB in double precision, far exceeding device memory.

\paragraph{Symmetry exploitation.}
The intermediate $X_{qs\mu}$ inherits the symmetry $X_{qs\mu}=X_{sq\mu}$ of the underlying
orbital pair $(q,s)$, because the gradient auxiliary basis enters only through the
symmetric combination $\sum_c\mathcal G^c_{\mu,g}\mathcal G^c_{\nu,g}$. The orbital
tiling therefore computes only the upper-triangular $(q\le s)$ panels and copies the
lower-triangular panels by transposition, cutting the work and storage footprint on
the non-factorizable terms roughly in half.

\paragraph{Out-of-core assembly.}
Two distinct multi-GPU paths are used. During ISDF kernel construction, the dense
grid dimension is sharded across devices; every device accumulates a partial
kernel, and a collective sum combines those partial results. After the kernels are
formed, four-index CCSD-facing orbital panels are independent: whole panels are
assigned round-robin to individual GPUs and written through a bounded asynchronous
queue, with no cross-device reduction between panels. Chunk-aligned on-disk
datasets and the two execution paths are detailed in Supporting Information
Section~\ref{si-sec:si-pipeline}.

\paragraph{Memory-aware tile selection.}
Given a user-supplied (or auto-detected) device-memory budget, the panel block size
$b_{\rm panel}$ is chosen by a binary search that fits the simultaneous peak
workspace: the ISDF intermediate $D$, the orbital tile of $X$, the pivot
tensor $C$, and the current output tile. The same procedure selects the grid-batch
size for the kernel computations so that the pipeline remains compute-bound on the
device.

\section{Computational Details}

All calculations were carried out with \textsc{pytc} release \texttt{v0.1.0}
(commit \texttt{09e0235}), interfaced with
\textsc{PySCF}\cite{pyscf2018,pyscf2020}; code and reproducibility scripts are
linked in the Data and Code Availability statement. The transcorrelated integral construction
and the coupled-cluster solver are implemented in JAX\cite{jax2018github} in double precision
throughout. The mean-field reference is restricted Hartree--Fock.

\paragraph{Grid and ISDF.} The production calculations use Becke partitioning,
Treutler--Ahlrichs radial nodes, Lebedev angular grids, and \textsc{PySCF}
level~2 pruning, a grid setting validated for transcorrelated integral evaluation
in Ref.~\onlinecite{haupt2023}, whose convergence tests we adopt here.
Interpolation points are obtained by pivoted Cholesky decomposition of the scalar
($\phi\phi$) and gradient ($\phi\nabla\phi$) densities. We denote the prescribed
rank of each channel by $N_{\rm ch}$, setting
$N_\mu^{(\phi)}=N_\mu^{(\nabla)}\equiv N_{\rm ch}=12\,N_{\rm orb}$, and take the
union of the two pivot sets. The final auxiliary dimension $N_\mu\equiv
N_{\rm fused}$ is listed in Table~\ref{tab:systems} and satisfies
$N_{\rm fused}\approx 1.8\,$--$\,2.2\,N_{\rm ch}$. The adequacy of this rank is
established in Fig.~\ref{fig:isdf_rank}.

\paragraph{Jastrow and VMC.} The correlator is a composite of a one-body
nuclear-cusp factor and a Boys--Handy form with $17$ terms per nucleus type
[Eqs.~(\ref{eq:bh-scaled})--(\ref{eq:bh-u})], yielding on the order of $20$ free
parameters for the hydrogen chain (one nucleus type) and $40$ for the molecular
systems (two nucleus types). The parameters are optimized by reference-variance
VMC as described in Section~\ref{sec:vmc}; a warm-up/refinement protocol is used,
and the production correlator is obtained by averaging the stationary tail of the
refinement trajectory. Walker counts, step counts, learning rates, burn-in, and
averaging-window details are collected in Supporting Information
Section~\ref{si-sec:si-vmc-schedule}.

\paragraph{Coupled cluster.} The non-Hermitian transcorrelated integrals are
passed to a restricted CCSD solver following the projective non-Hermitian
transcorrelated coupled-cluster framework of
Refs.~\onlinecite{liao2021e,schraivogel2021,schraivogel2023}
(maximum $100$ iterations, level shift $0.1$), with the $(vvvv)$ block contracted
on the fly. The rate-limiting $(vvvv)$ integral tiles feeding the $\hat T_2$
residual are assembled on the GPU through the same multi-device pipeline used for
the ISDF construction, while the remaining amplitude-update contractions run on the
host. For the largest benzene calculation (cc-pCV5Z, $N_{\rm orb}=1200$) the CCSD
iterations ran on eight NVIDIA B200 GPUs and reached practical convergence in
${\sim}29$~h.

\paragraph{Systems and hardware.} Table~\ref{tab:systems} collects the system
sizes and ISDF ranks. The hydrogen-chain rows report the cc-pV5Z MP2NO
production spaces used for the large-chain calculations and cutoff checks; the
energy figures below retain explicit basis labels. Ethylene uses cc-pVTZ,
and benzene uses the cc-pCVDZ, cc-pCVTZ, cc-pCVQZ, and cc-pCV5Z bases.
Calculations were run on NVIDIA A100, H200, and B200 GPUs according to system size:
the H$_{10}$ and small molecular references used a single A100 or H200, the longer
hydrogen chains (H$_{20}$--H$_{50}$) used single H200 or B200 devices, and the
benzene cc-pCV5Z calculation used eight B200 GPUs.

\begin{table}[t]
    \caption{System sizes and ISDF parameters: number of orbitals $N_{\rm orb}$,
    grid points $N_g$ (\textsc{PySCF} level~2), and fused auxiliary dimension
    $N_{\rm fused}$ (union of the scalar and gradient pivot sets, each capped at
    $N_{\rm ch} = 12\,N_{\rm orb}$). For the hydrogen-chain rows, the listed
    cc-pV5Z entries are production MP2NO orbital spaces: $N_{\rm orb}$ is the
    occupied Hartree--Fock space plus the retained virtual MP2 natural orbitals, and
    $N_{\rm fused}$ is the corresponding ISDF fused dimension.}
    \label{tab:systems}
    \centering
    \begin{tabular}{llrrr}
        \hline\hline
        System & Basis & $N_{\rm orb}$ & $N_g$ & $N_{\rm fused}$ \\ \hline
        H$_{10}$  & cc-pV5Z   & 105  & 53\,720  & 2\,363 \\
        H$_{20}$  & cc-pV5Z   & 210  & 107\,440 & 4\,547 \\
        H$_{30}$  & cc-pV5Z   & 315  & 161\,160 & 6\,833 \\
        H$_{40}$  & cc-pV5Z   & 420  & 214\,880 & 9\,077 \\
        H$_{50}$  & cc-pV5Z   & 525  & 268\,600 & 11\,592 \\
        C$_2$H$_4$ & cc-pVTZ  & 116  & 43\,904  & 2\,460 \\
        C$_6$H$_6$ & cc-pCVDZ & 138  & 99\,480 & 3\,670 \\
        C$_6$H$_6$ & cc-pCVTZ & 342  & 99\,480 & 8\,877 \\
        C$_6$H$_6$ & cc-pCVQZ & 684  & 99\,480 & 16\,389 \\
        C$_6$H$_6$ & cc-pCV5Z & 1200 & 99\,480 & 25\,894 \\
        \hline\hline
    \end{tabular}
\end{table}

\section{Results}

\subsection{Accuracy of the ISDF approximation}
\begin{figure}[htbp]
    \centering
    \includegraphics[width=\columnwidth]{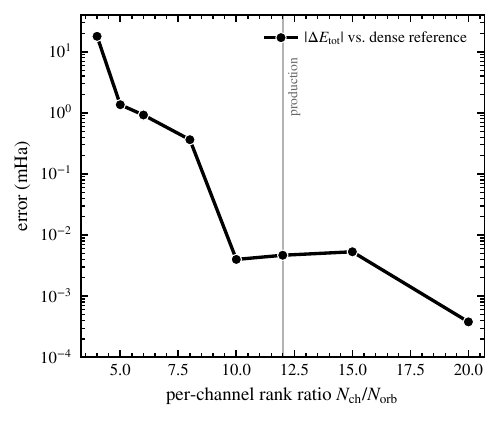}
    \caption{Convergence of the ISDF approximation with rank ratio
    $N_{\rm ch}/N_{\rm orb}$ for ethylene (C$_2$H$_4$, cc-pVTZ, $N_{\rm orb}=116$,
    VMC-optimized Boys--Handy Jastrow): ISDF-xTC-CCSD total-energy error
    $|\Delta E_{\rm tot}|$ measured relative to an explicit dense non-ISDF
    reference. The vertical line marks the production setting
    $N_{\rm ch} = 12\,N_{\rm orb}$.}
    \label{fig:isdf_rank}
\end{figure}

Because the entire scheme rests on a low-rank ISDF representation of the
transcorrelated integrals, we first verify that this approximation is
systematically convergent. We do not evaluate the full three-body $L$ tensor in
the production calculations; the approximation being accelerated is therefore
the established xTC Hamiltonian rather than a dense three-body transcorrelated
reference. Prior work has benchmarked xTC and non-Hermitian transcorrelated
solvers in molecular and strongly correlated settings, including
transcorrelated N$_2$ binding and dissociation
curves\cite{christlmaier2023,schraivogel2023,haupt2025,liao2023}; the
convergence test here therefore isolates the additional ISDF truncation error
introduced by the grid-based factorization. For ethylene (C$_2$H$_4$, cc-pVTZ,
$N_{\rm orb}=116$) with a VMC-optimized Boys--Handy Jastrow,
Fig.~\ref{fig:isdf_rank} reports the ISDF-xTC-CCSD total-energy error as a
function of the per-channel rank ratio $N_{\rm ch}/N_{\rm orb}$, measured against an explicit
dense (non-ISDF) transcorrelated reference evaluated with the same Jastrow and
solver convention. The dense reference gives
$E_{\rm tot}=-78.4904432130$~Ha, while the highest ISDF rank shown
($N_{\rm ch} = 20\,N_{\rm orb}$) gives $E_{\rm tot}=-78.4904435930$~Ha, differing by
only $3.8\times10^{-7}$~Ha ($0.38~\mu$Ha). The ISDF truncation error in the
total energy falls below $1$~mHa by $N_{\rm ch} \approx 6\,N_{\rm orb}$ and is
converged to $<0.01$~mHa by $N_{\rm ch} = 10\,N_{\rm orb}$. The lowest ratio shown
($N_{\rm ch} = 4\,N_{\rm orb}$) is under-resolved (error ${\sim}18$~mHa) because the
more diffuse triple-zeta basis raises the numerical rank of the orbital-product
densities. Since the ISDF pivots are selected from the orbital products alone
(Alg.~\ref{alg:pivot}), independently of the correlator, alternative
differentiable correlators can be used without changing the pivot-selection
procedure. The rank-convergence benchmark shown here uses the Boys--Handy/nuclear-cusp
correlator of the production calculations.

\subsection{Hydrogen chain in the thermodynamic limit}
\begin{figure}[htbp]
    \centering
    \includegraphics[width=\columnwidth]{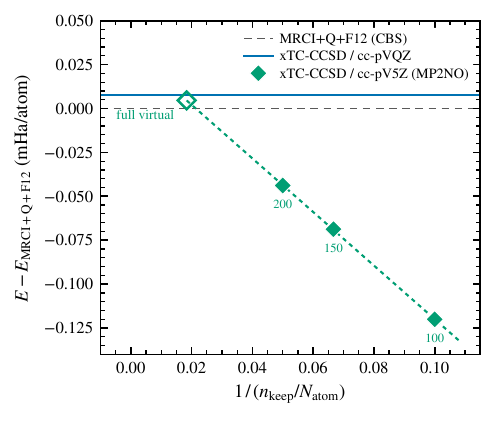}
    \caption{Convergence of the H$_{10}$ cc-pV5Z MP2NO virtual-orbital cutoff at
    $r=1.4$~bohr. The horizontal axis is the inverse retained-orbital count per
    atom, $1/(n_{\mathrm{keep}}/N_{\mathrm{atom}})$, so the full-virtual limit
    lies at the finite abscissa $1/(n_{\rm vir}/N_{\mathrm{atom}})=10/544$
    (cc-pV5Z has $n_{\rm vir}=544$ for H$_{10}$); the vertical axis is the
    per-atom energy deviation from
    the MRCI+Q+F12 reference. Filled diamonds are the finite
    $n_{\mathrm{keep}}/N_{\mathrm{atom}}=10$, $15$, $20$ results; the dashed line
    is an unweighted linear least-squares fit, evaluated at the full-virtual
    abscissa (open diamond) to estimate the full-virtual limit. The dashed
    horizontal line marks the
    reference and the solid line the full-virtual cc-pVQZ result. The
    extrapolated cc-pV5Z limit agrees with cc-pVQZ to ${\sim}0.003$~mHa/atom,
    indicating that cc-pVQZ is essentially at the basis-set limit here, and
    the production cutoff $n_{\mathrm{keep}}/N_{\mathrm{atom}}=10$ lies within
    ${\sim}0.12$~mHa/atom of the limit, supporting its use for the longer
    chains.}
    \label{fig:h10_5z_nkeep}
\end{figure}

\begin{figure}[htbp]
    \centering
    \includegraphics[width=\columnwidth]{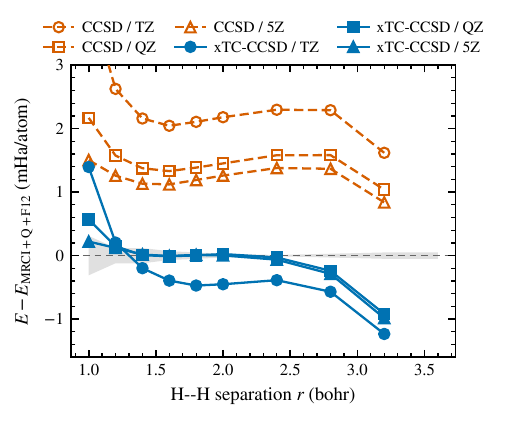}
    \caption{Basis-set convergence on the H$_{10}$ chain: per-atom energy
    deviation from the MRCI+Q+F12 CBS-limit reference for xTC-CCSD (filled symbols) and
    canonical CCSD (open symbols) at cc-pVTZ (circles), cc-pVQZ (squares), and
    cc-pV5Z (triangles), versus H--H separation. For xTC-CCSD the cc-pV5Z point is
    the MP2NO series ($n_{\rm keep}=100,200,300$) extrapolated to the finite
    full-virtual limit $n_{\rm keep}=n_{\rm vir}=544$; for canonical CCSD it is the full (untruncated)
    cc-pV5Z result. Both therefore reach the full-virtual limit. The shaded band is the
    reference extrapolation
    uncertainty. xTC-CCSD at cc-pVTZ already outperforms canonical CCSD at
    cc-pV5Z, and its cc-pVQZ and cc-pV5Z results coincide within the reference uncertainty
    over the bound region.}
    \label{fig:h10_basis}
\end{figure}

For the cc-pV5Z production orbital spaces summarized in Table~\ref{tab:systems},
we use MP2 natural-orbital (MP2NO) truncations: the virtual space is rotated to the
eigenbasis of the MP2 one-particle density matrix and truncated by natural-orbital
occupation\cite{dobrautz2024}, retaining
$n_{\mathrm{keep}}/N_{\mathrm{atom}}=10$ virtual natural orbitals per atom.
A complementary route to reducing basis-set errors at small-basis cost is the
additive reference correction of Hauskrecht \emph{et al.}\cite{hauskrecht2026},
which retains the small-basis xTC correlation energy while importing the
reference contribution from a larger basis; here we instead work directly in
the large basis with MP2NO-truncated virtual spaces.
We first establish the accuracy of this truncation and how the cc-pV5Z data
are obtained. Figure~\ref{fig:h10_5z_nkeep}
shows the cc-pV5Z convergence test for H$_{10}$ at $r=1.4$~bohr, a representative
short-side bound-region separation. This focused cutoff diagnostic uses
$n_{\rm keep}=100,150,200$, whereas the binding-curve extrapolations in
Fig.~\ref{fig:h10_basis} below use $100,200,300$ at every separation; the two datasets
serve different purposes and are fitted independently using the same inverse-count
form. Increasing the cutoff from 10 to 15 and
20 retained virtuals per atom changes the energy by only $0.051$ and
$0.025$~mHa/atom, respectively. Because cc-pV5Z has a finite virtual space
($n_{\rm vir}=544$ for H$_{10}$), the physical full-virtual limit lies at the
finite abscissa $1/(n_{\rm vir}/N_{\mathrm{atom}})=10/544\approx0.018$ rather
than at the origin; evaluating the linear fit in
$1/(n_{\mathrm{keep}}/N_{\mathrm{atom}})$ there places the full-virtual
cc-pV5Z limit within $0.005$~mHa/atom of the MRCI+Q+F12 reference,
essentially coincident with the full-virtual cc-pVQZ result
($+0.008$~mHa/atom). The extrapolated 5Z and QZ full-virtual energies thus
agree to ${\sim}0.003$~mHa/atom, indicating that cc-pVQZ is already
essentially at the basis-set limit for this system. The finite
$n_{\mathrm{keep}}/N_{\mathrm{atom}}=10$ production cutoff is within about
$0.12$~mHa/atom of that extrapolated limit, well below the $1$~mHa/atom scale
used in the chain benchmarks.
Because only three cutoff points enter this empirical inverse-count fit, its
intercept has no independent statistical error estimate and can retain fit-form
bias; the quoted sub-mHa comparisons diagnose the selected linear extrapolation
rather than establish sub-mHa certainty in the full-virtual limit.

With the MP2NO truncation validated, Fig.~\ref{fig:h10_basis} compares
xTC-CCSD and canonical
CCSD at the cc-pVTZ, cc-pVQZ, and cc-pV5Z levels against an accurate MRCI+Q+F12
reference~\cite{motta2017}; for xTC-CCSD the cc-pV5Z curve is obtained from the
MP2NO series exactly as established above.
Canonical CCSD remains well above the reference at every basis---$1.3$--$2.3$~mHa/atom
at cc-pVQZ and still ${\sim}1.1$--$1.5$~mHa/atom even at the full cc-pV5Z basis---whereas
xTC-CCSD lies within $0.6$~mHa/atom at cc-pVTZ over $r=1.2$--$2.8$~bohr
and, over the central bound region $r=1.4$--$2.4$~bohr, agrees with the reference to within $0.1$~mHa/atom
at cc-pVQZ---comparable to the reference's own extrapolation uncertainty. Over the broader $1.2$--$2.8$~bohr window,
xTC-CCSD at cc-pVTZ already surpasses canonical CCSD at cc-pV5Z. This rapid convergence motivates retaining
the cc-pVTZ thermodynamic-limit series as a lower-cost reference trend, with
the residual basis and virtual-orbital cutoff error quantified by the cc-pV5Z
MP2NO diagnostic above.

\FloatBarrier

\begin{figure}[htbp]
    \centering
    \includegraphics[width=\columnwidth]{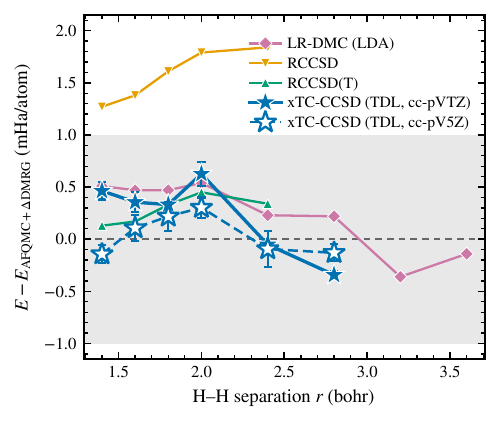}
    \caption{Deviation of the per-atom energy from the AFQMC+$\Delta$DMRG
    reference for the hydrogen chain as a function of H--H separation $r$.
    The ISDF-xTC-CCSD thermodynamic-limit result (this work, cc-pVTZ; filled stars) stays within
    $\pm1$~mHa/atom (shaded band) across the studied xTC range
    of separations (1.4--2.8~bohr), outperforming RCCSD and tracking
    CBS-extrapolated RCCSD(T) and continuous-space LR-DMC(LDA).
    Open stars (dashed) show the cc-pV5Z MP2NO thermodynamic-limit series, which
    reduces the maximum absolute reference deviation from $0.63$ to
    $0.31$~mHa/atom and supports residual orbital-basis effects below about
    $0.6$~mHa/atom over this range at the stated MP2NO cutoff. Error bars on
    both TDL series show the $1\sigma$ intercept standard errors of the
    $1/N$ fits (Tables in the Supporting Information).}
    \label{fig:tdl_dev}
\end{figure}

We benchmark ISDF-xTC-CCSD on the linear hydrogen chain, a canonical
strongly-correlated model, over a range of interatomic separations
$r = 1.4$--$2.8$~bohr. The upper end of this range reflects the domain where the
single-reference, non-Hermitian CCSD solver converges robustly for these
transcorrelated chains; more stretched geometries require a multireference
treatment beyond the present solver. For the TDL series plotted here, each $r$
uses finite chains of $N = 20$, $30$, $40$, and $50$ atoms in the cc-pVTZ basis;
we extrapolate the total energy per atom to the thermodynamic limit (TDL) by a linear
least-squares fit in $1/N$, $E_{\rm tot}(N)/N = E_{\rm TDL} + a/N$
(the per-separation fits are collected in the Supporting Information,
Fig.~\ref{si-fig:tdl_extrap}). Residual and fit-form sensitivity checks are reported
there; they support the headline $1$~mHa/atom comparison while cautioning against
overinterpreting differences of a few tenths of a mHa.
The extrapolated TDL energies are compared with
the AFQMC+$\Delta$DMRG reference data of Motta \emph{et al.}~\cite{motta2017},
alongside the LR-DMC(LDA), RCCSD, and RCCSD(T) results from the same
compilation\cite{motta2017} (Fig.~\ref{fig:tdl_dev}). Across
all separations studied, ISDF-xTC-CCSD agrees with the reference to within
$\pm 1$~mHa/atom, a substantial improvement over canonical RCCSD, whose deviation
approaches $2$~mHa/atom near equilibrium. It matches the accuracy of RCCSD(T) and
LR-DMC(LDA) while retaining the $\mathcal{O}(N^{6})$ cost of CCSD, avoiding both the
$\mathcal{O}(N^{7})$ perturbative-triples step of RCCSD(T) and the stochastic
sampling of the diffusion Monte Carlo references.
An analogous TDL series in the cc-pV5Z basis with the MP2NO cutoff
$n_{\mathrm{keep}}/N_{\mathrm{atom}}=10$ (open stars in Fig.~\ref{fig:tdl_dev};
per-separation fits in Fig.~\ref{si-fig:tdl_extrap_5z} and
Table~\ref{si-tab:si-tdl-5z} of the Supporting Information) reduces the maximum
absolute reference deviation from $0.63$ to $0.31$~mHa/atom and supports
residual orbital-basis effects below about $0.6$~mHa/atom over this range at
the stated MP2NO cutoff.

\subsection{Ground-state energy of benzene}

\begin{figure}[htbp]
    \centering
    \includegraphics[width=\columnwidth]{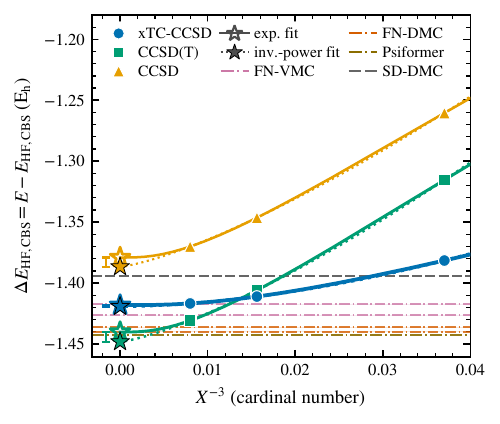}
    \caption{Complete-basis-set extrapolation of the benzene energy relative to
    the CBS Hartree--Fock limit, $\Delta E_{\rm HF,CBS}=E-E_{\rm HF,CBS}$, plotted against $X^{-3}$ (cardinal number $X$ of
    cc-pCV$X$Z, $X=\mathrm{D,T,Q,5}$). The horizontal axis is used only as a conventional visualization
    of cardinal-number convergence; the xTC-CBS values are obtained from
    exponential and inverse-power extrapolations, not from an assumed $X^{-3}$ law.
    ISDF-xTC-CCSD (this work) is compared with canonical CCSD and CCSD(T)~\cite{ren2023}.
    Horizontal lines show literature benchmarks from FermiNet-VMC/DMC~\cite{ren2023},
    Psiformer~\cite{vonglehn2023}, and best single-determinant DMC. Solid and dotted curves denote the two CBS fits;
    open and filled stars mark the corresponding extrapolated CBS values. Error
    bars at the CBS limit span the exponential-versus-inverse-power fit-form
    spread ($1.1$, $7.6$, and $8.0$~mHa for xTC-CCSD, CCSD, and CCSD(T)); with
    exact three-point fits this is a fit-form sensitivity, not a statistical
    standard error.
    Numerical values are collected in Supporting Information
    Table~\ref{si-tab:si-benzene-cbs}.}
    \label{fig:benzene_corr}
\end{figure}

Figure~\ref{fig:benzene_corr} shows the basis-set convergence of the benzene
energy relative to the CBS Hartree--Fock limit,
$\Delta E_{\rm HF,CBS}=E-E_{\rm HF,CBS}$, across the cc-pCV$X$Z series. At finite
$X$ this quantity is not the conventional basis-specific correlation energy
$E(X)-E_{\rm HF}(X)$; only its CBS limit equals the CBS correlation energy.
Extrapolating to the complete-basis-set (CBS) limit, ISDF-xTC-CCSD recovers
$E_{\rm corr,CBS} \approx -1.418$~$E_{\rm h}$, which lies between the canonical
CCSD ($\approx -1.379$~$E_{\rm h}$) and CCSD(T) ($\approx -1.440$~$E_{\rm h}$)
CBS values~\cite{ren2023}: about $39$~mHa more correlation than CCSD, but still
roughly $22$~mHa short of the CCSD(T)/CBS benchmark. It closely matches the
three-layer FermiNet variational Monte Carlo (VMC) result
($\approx -1.417$~$E_{\rm h}$)~\cite{ren2023}; the deeper four-layer VMC
($\approx -1.426$~$E_{\rm h}$) and the fixed-node diffusion Monte Carlo (DMC)
FermiNet results ($\approx -1.436$ and $-1.440$~$E_{\rm h}$ for the three- and
four-layer ans\"atze) recover progressively more correlation, the four-layer
FermiNet-DMC essentially reaching the CCSD(T)/CBS benchmark~\cite{ren2023}.
Since xTC introduces no explicit
triples, this CCSD-level method captures about $64\%$ of the CCSD-to-CCSD(T) CBS
gap ($39$ of $61$~mHa) at the formal scaling of CCSD, without a
perturbative-triples step or explicitly correlated neural-network ans\"atze such
as the Psiformer~\cite{vonglehn2023}.
Beyond the CBS value itself, transcorrelation makes the extrapolation markedly
more robust to the choice of fit form: the exponential and inverse-power
extrapolations of the xTC-CCSD correlation energy agree to $1.1$~mHa, whereas for
canonical CCSD and CCSD(T) they differ by $7.6$ and $8.0$~mHa, respectively. Each CBS value is a three-parameter fit to the three points (cc-pCVTZ/QZ/5Z), so this spread quantifies the sensitivity to the fit form, not a statistical uncertainty; the fit forms and fitted parameters are tabulated in the Supporting Information. The
smoother, accelerated basis-set convergence of the transcorrelated correlation
energy renders the extrapolated CBS limit far less sensitive to the extrapolation
scheme.

\begin{figure}[htbp]
    \centering
    \includegraphics[width=\columnwidth]{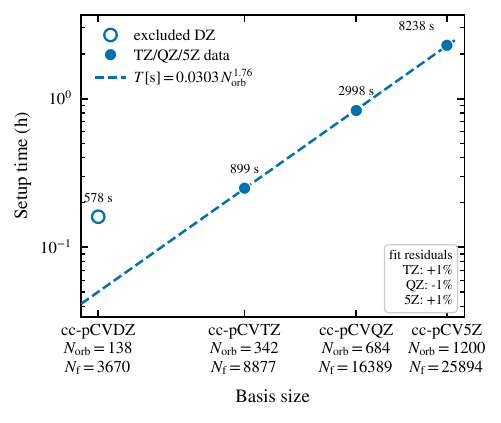}
    \caption{Basis-set scaling of the ISDF-xTC setup time for benzene on a
    single B200 GPU. Filled points (cc-pCVTZ/QZ/5Z) define the power-law fit
    $T\,[\mathrm{s}] = 0.030\,N_{\rm orb}^{1.76}$ (dashed); the cc-pCVDZ point (open) is
    excluded from the fit. Annotations give the absolute setup time in
    seconds; the lower axis lists $N_{\rm orb}$ and the fused auxiliary dimension
    $N_{\rm f}\equiv N_{\rm fused}$. Setup covers grid/orbital preparation,
    ISDF decomposition, and $K$/$\Delta U$ kernel/intermediate construction, but
    excludes SCF/VMC, four-index materialization, and CCSD iterations.}
    \label{fig:benzene_timing}
\end{figure}

Figure~\ref{fig:benzene_timing} reports the wall-clock time required to build
the ISDF-xTC intermediates for benzene on a single B200 GPU
across the cc-pCV$X$Z ($X = $ D, T, Q, 5) series, spanning
$N_{\rm orb} = 138$ to $1200$. The setup time scales empirically as
$T\,[\mathrm{s}] \approx 0.030\,N_{\rm orb}^{1.76}$ over this range, below the formal
$\mathcal{O}(N_{\rm orb}^{2}N_\mu^{2})$ worst case; we emphasize that this is a
three-point power-law fit over the accessible range rather than an asymptotic
scaling result, with the largest
cc-pCV5Z calculation ($N_{\rm orb} = 1200$, $N_{\rm fused} = 25894$) completing in
$\approx 2.3$~h. Here ``setup'' starts with grid/orbital preparation and includes the fixed-rank ISDF decomposition and the
$K$/$\Delta U$ kernel/intermediate construction.
It excludes SCF, VMC optimization and history loading, four-index CCSD-facing
block materialization, and the CCSD iterations.

\section{Conclusions}

We have presented an ISDF construction of the transcorrelated Hamiltonian for
general differentiable Jastrow correlators. The method evaluates the
transcorrelated integrals on a real-space grid, compresses the orbital-pair and
orbital-gradient densities into a low-rank ISDF form, and combines this with the
xTC reduction of the three-body operator to an effective two-body correction.
Because the compact auxiliary dimension $N_\mu$ (a small multiple of
$N_{\rm orb}$) replaces both the dense grid index $N_g$ in the final integral
tensors and the orbital-pair count $N_{\rm orb}^{2}$ in the kernel intermediates,
this lowers grid-dependent storage and integration costs by orders of magnitude
relative to dense grid evaluation while retaining the flexibility of numerical
integration. Automatic differentiation supplies Jastrow gradients, and an
out-of-core GPU producer/consumer pipeline makes the construction tractable for
hundreds of orbitals and large basis sets.

The ISDF approximation is systematically convergent: for ethylene in a
triple-zeta basis the ISDF truncation error in the total energy falls well below
$1$~mHa by a per-channel rank ratio $N_{\rm ch}/N_{\rm orb}\approx 10$---with the production
setting $N_{\rm ch}=12\,N_{\rm orb}$ comfortably converged. On the linear
hydrogen chain, transcorrelation strongly accelerates basis-set convergence:
xTC-CCSD at the triple-zeta level already surpasses canonical CCSD at
quintuple-zeta over $r=1.2$--$2.8$~bohr, and the thermodynamic-limit energies agree with
AFQMC+$\Delta$DMRG references to within $1$~mHa/atom across a wide range of
bond lengths. For benzene, the method recovers a complete-basis-set correlation
energy well beyond that of canonical CCSD and comparable to neural-network
variational Monte Carlo, while remaining tractable up to cc-pCV5Z. In both cases
transcorrelation substantially improves the accuracy of an otherwise approximate
coupled-cluster treatment, recovering---at the formal scaling of CCSD---a substantial part of the correlation energy that
conventionally requires perturbative triples or a considerably larger basis.

Natural next steps include periodic systems\cite{simula2025solids},
pseudopotentials for heavier elements\cite{simula2025pp}, higher-order
coupled-cluster treatments of the transcorrelated Hamiltonian\cite{morchen2025},
stochastic treatments (e.g.\ AFQMC), and alternative or neural-network Jastrow forms.
More broadly, the combination of automatically differentiable correlators with
low-rank, GPU-accelerated integral construction makes the transcorrelated method a
practical route to high accuracy in compact basis sets across a range of
downstream solvers, from coupled cluster to stochastic and tensor-network
approaches.

\section*{Data and Code Availability}
The numerical data plotted in all figures, together with the plotting scripts
that reproduce them, are available in the repository
\url{https://github.com/nickirk/tc-isdf-data}.
The \textsc{pytc} code used to produce these results is available at
\url{https://github.com/nickirk/pytc} (release \texttt{v0.1.0}, commit
\texttt{09e0235}; also distributed as \texttt{pytc-qc}~0.1.0 on PyPI).

\section*{Acknowledgments.}
K.L. (the work performed at Yale) and T.Z. are supported by the National Science Foundation under Grant No. CHE-2337991.
K.L. (the work performed at Max Planck Institute for Solid State Research) is supported by the Max Planck Society.
W. D. acknowledges funding from the German Federal Ministry of Research, Technology and Space (BMFTR) under the research program Quantensysteme and funding measure Quantum Futur 3 for project No. 13N17229 as well as the Helmholtz Association via Initiative and Networking Fund (INF) for projects No. VH-NG-21-08 under the Helmholtz Investigator program as well Projects KA-QUS-02 (qFLOW) and KA-QUS-03 (QT-Batt) under the Helmholtz Quantum Use case call. A.A. acknowledges generous funding from the Max Planck Society. 
The authors thank the Yale Center for Research Computing
for providing computational resources and support that contributed to the results reported within this paper.

\bibliography{tc-isdf}

\end{document}